\documentclass[
reprint,
superscriptaddress,
amsmath,
amssymb,
aps,
pra,
twocloumn]{revtex4-2}
\usepackage{graphicx}
\usepackage{dcolumn}
\usepackage{bm}
\usepackage{physics}
\usepackage{appendix}
\usepackage[normalem]{ulem}
\usepackage{mathrsfs}
\usepackage{comment}
\usepackage{subfigure}
\usepackage{svg}
\usepackage{todonotes}
\usepackage{csquotes}
\usepackage{bbm}
\usepackage{dsfont}
\usepackage{hyperref}
\usepackage[capitalise]{cleveref}

\makeatletter
\pdfpageheight\paperheight
\pdfpagewidth\paperwidth

\makeatother
\newcommand{\ulm}{Institute for Complex Quantum Systems, Ulm University, 89069 Ulm, Germany}
\newcommand{\iqst}{Center for Integrated Quantum Science and Technology (IQST), Ulm-Stuttgart, Germany}

\begin{document}
\title{Boosting thermalization of classical and quantum many-body systems}
\author{Jin-Fu Chen}
\email{jinfuchen@lorentz.leidenuniv.nl}
\affiliation{Instituut-Lorentz, Universiteit Leiden, P.O. Box 9506, 2300 RA Leiden, The Netherlands}
\affiliation{$\langle aQa^L\rangle$ Applied Quantum Algorithms Leiden, The Netherlands}
\author{Kshiti Sneh Rai}
\affiliation{Instituut-Lorentz, Universiteit Leiden, P.O. Box 9506, 2300 RA Leiden, The Netherlands}
\affiliation{$\langle aQa^L\rangle$ Applied Quantum Algorithms Leiden, The Netherlands}
\author{Patrick Emonts}
\affiliation{Instituut-Lorentz, Universiteit Leiden, P.O. Box 9506, 2300 RA Leiden, The Netherlands}
\affiliation{$\langle aQa^L\rangle$ Applied Quantum Algorithms Leiden, The Netherlands}
\affiliation{\ulm}
\affiliation{\iqst}
\author{Donato Farina}
\affiliation{Physics Department E. Pancini - Università degli Studi di Napoli Federico II, Complesso Universitario Monte S. Angelo - Via Cintia - I-80126 Napoli, Italy.
}
\affiliation{ICFO-Institut de Ciencies Fotoniques, The Barcelona Institute of Science and Technology, Castelldefels (Barcelona) 08860, Spain}
\author{Marcin Płodzień}
\affiliation{ICFO-Institut de Ciencies Fotoniques, The Barcelona Institute of
Science and Technology, Castelldefels (Barcelona) 08860, Spain}
\author{Przemyslaw Grzybowski}
\affiliation{ICFO-Institut de Ciencies Fotoniques, The Barcelona Institute of
Science and Technology, Castelldefels (Barcelona) 08860, Spain}
\affiliation{Institute of Spintronics and Quantum Information, Faculty of Physics and Astronomy, Adam Mickiewicz University, Umultowska 85, 61-614 Pozna{\'n}, Poland}

\author{Maciej Lewenstein}
\affiliation{ICFO-Institut de Ciencies Fotoniques, The Barcelona Institute of
Science and Technology, Castelldefels (Barcelona) 08860, Spain}
\affiliation{ICREA – Institució Catalana de Recerca i Estudis Avançats, Lluís Companys 23, 08010 Barcelona, Spain}
\author{Jordi Tura}
\email{tura@lorentz.leidenuniv.nl}
\affiliation{Instituut-Lorentz, Universiteit Leiden, P.O. Box 9506, 2300 RA Leiden, The Netherlands}
\affiliation{$\langle aQa^L\rangle$ Applied Quantum Algorithms Leiden, The Netherlands}

\date{\today}
\begin{abstract}
Understanding and optimizing the relaxation dynamics of many-body systems is essential both for foundational studies in quantum thermodynamics and for applications such as quantum simulation and quantum computing. Efficient preparation of thermal states of a many-body Hamiltonian is governed by the spectral properties of the associated Lindbladian, in particular its spectral gap, which determines the slowest relaxation rate.
In this work, we develop a systematic framework for constructing Lindbladians that prepare thermal states. Our approach reveals a simple relation between the relaxation dynamics at finite and infinite temperatures. The framework is scalable to larger system sizes when implemented using tensor-network methods. We find that efficient thermalization requires that the relaxation dynamics respect the symmetries of the thermal state, which reduces the number of free parameters. By applying gradient-based optimization to the Lindbladians, we enhance the spectral gap and thereby boost thermalization. When applied to both classical and quantum spin models, our method demonstrates a substantial enhancement of the spectral gap. For larger system sizes, our approach provides a variational upper bound and enables a certified lower bound on the minimum relaxation rate.
\end{abstract}
\maketitle

\newpage
\section{Introduction}

Significant progress has been made in controlling quantum many-body systems using precise experimental manipulation techniques, which enable the construction of desired quantum states and the implementation of target Hamiltonians.
Experimental platforms such as ultracold atoms~\cite{leonard_realization_2023}, superconducting qubits~\cite{google_quantum_ai_suppressing_2023_short}, photons ~\cite{wang_integrated_2020,madsen_quantum_2022}, and trapped ions~\cite{kokail_self-verifying_2019} among others, offer near-term quantum devices for implementation. 
These advancements have enabled the exploration of nonequilibrium dynamics for quantum many-body systems~\cite{polkovnikov_colloquium_2011}, by designing complex interactions~\cite{periwal_programmable_2021} and control protocols~\cite{lukin_quantum_2024}. 
However, real quantum systems cannot be fully isolated from their environments, and the unitary evolution of closed systems remains an approximation.

The next developmental leap in the field of quantum simulators requires controlling dissipation and decoherence dynamics by engineering of Lindbladians, in order to realize the target properties of open system dynamics.
In particular, preserving quantum properties over dissipation dynamics is crucial from a practical point of view.
In quantum computing, addressing dissipative effects is essential for maintaining coherence and minimizing errors, which are critical for reliable operation~\cite{li_efficient_2017,temme_error_2017}.
One prominent example is dissipative quantum computing, where controlled interactions with the environment are utilized to stabilize quantum states and facilitate computations \cite{verstraete_quantum_2009,lin_dissipative_2013,reiter_dissipative_2017,harrington_engineered_2022}.
Additionally, dissipation can be exploited to engineer non-trivial quantum phases, such as topological states of matter \cite{Diehl2011,PhysRevB.101.125412,PhysRevLett.127.245701,PhysRevResearch.4.023036}, and to induce inherently quantum correlations, like entanglement \cite{Kordas_2012,PhysRevB.104.184422,10.21468/SciPostPhys.12.1.011}. This engineered relaxation dynamics demonstrates that environmental noise can be exploited as a resource in quantum technologies, rather than merely being an obstacle to overcome.

Thermal states represent the typical steady state after relaxation, but preparing quantum thermal states at low temperature is believed to be intractable, even for quantum computers~\cite{alhambra_quantum_2023}. 
At higher temperatures, however, thermal-state preparation becomes more tractable due to reduced correlations and entanglement.
Nevertheless, efficient quantum thermal state preparation \cite{ge_rapid_2016,chen2023quantum, PhysRevA.110.012445} is useful for quantum simulation and quantum computing, enabling exploring phases and phase transitions in quantum many-body systems \cite{guo_new_2024} and enhancing other quantum algorithms \cite{holmes_quantum_2022,Rouze2024}, with broad applications in quantum simulation, optimization, and open quantum systems. 
Notably, Monte Carlo-style quantum Gibbs samplers \cite{temme2011quantum} have emerged as a cornerstone.
Recent studies have developed efficient quantum Gibbs samplers for noncommuting Hamiltonians~\cite{chen_efficient_2025, gilyen_quantum_2024,ding_single-ancilla_2024}, implemented by quasilocal Lindbladians that exactly satisfy the quantum detailed balance condition ~\cite{fagnola_generators_2010,ramezani_quantum_2018}. The operators in Lindbladians are mainly localized within a region, with temperature-dependent decaying tails extending outside.

In recent years there has been widespread interest in studying the spectral properties of Lindbladians in open quantum many-body systems \cite{PhysRevLett.123.140403,PhysRevA.100.062131,amato2024number, muratore2024universal, Mori2020},  with particular attention on the spectral gap \cite{ PhysRevLett.130.230404, PhysRevB.109.064311,kastoryano_quantum_2016,bardet_rapid_2023,alicki_thermalization_2009,PhysRevLett.126.160401}. Solving and engineering the Lindbladian spectrum enables characterizing the relevant relaxation timescales in many-body systems \cite{PhysRevLett.111.150403, PhysRevLett.125.230604, PhysRevLett.127.070402}, and also paving the way for novel quantum simulation paradigms \cite{PhysRevX.6.041031}, classifications of open quantum criticality \cite{PhysRevLett.129.120401,minganti_spectral_2018}, explorations of integrability to chaos transitions  \cite{rubio2022chaos,qiao_quantum_2025}.

\subsection{Summary of results}

We introduce a transformation that connects finite-temperature Lindbladians to their infinite-temperature counterparts. This construction clarifies the role of recently proposed exact quantum Gibbs samplers for noncommuting Hamiltonians \cite{chen_efficient_2025, gilyen_quantum_2024, ding_single-ancilla_2024}. For quantum many-body systems, our approach exploits the dynamical Lie algebra and can be implemented with the tensor-network methods after truncating the Lindbladians to moderately local terms, thereby enabling scalability with system size. For classical systems, the same transformation naturally gives Liouvillians for preparing classical thermal states. In both cases, we construct Lindbladians (or Liouvillians in the classical case) from quasilocal (local) operators to ensure implementability on near-term devices.

Under the quantum detailed balance condition~\cite{fagnola_generators_2010,ramezani_quantum_2018}, we map the problem of computing the spectral gap of a Lindbladian to that of evaluating the energy gap of a frustration-free parent Hamiltonian in an extended Hilbert space. The spectral gap can be bounded from above with variational approaches such as the density-matrix renormalization group (DMRG)~\cite{white_density_1992,schollwock_density-matrix_2011,stoudenmire_studying_2012}. 
However, an upper bound alone is insufficient to certify the minimum relaxation rate; certified lower bounds are also required and can be obtained via semidefinite programming (SDP)~\cite{cruz_preparation_2022,rai_hierarchy_2024,kull_lower_2024,robichon_bootstrapping_2024,mortimer_certifying_2025}. Since the parent Hamiltonian and the Lindbladian share the same spectrum, this approach directly certifies relaxation rates in open many-body systems.

We apply gradient-based optimization to a family of Lindbladians to enhance the spectral gap. By mapping Lindbladians to parent Hamiltonians, we extend spectral-gap optimization techniques~\cite{giudici_locality_2022,rai_spectral_2025} to open quantum systems, providing a variational framework for optimizing relaxation dynamics under the detailed balance condition. We find that, to boost thermalization, the kinetic rules should obey the symmetries of the thermal state, thereby enhancing the spectral gap and reducing the number of free parameters in the Lindbladian. Optimizing the spectral gap facilitates efficient control of irreversibility and fluctuations in open many-body systems, since rapid thermalization relative to the control timescale is essential for implementing optimal slow-driving protocols in quantum and stochastic thermodynamics~\cite{rezakhani_quantum_2009,sivak_thermodynamic_2012,PhysRevLett.119.050601,scandi_thermodynamic_2019,chen_extrapolating_2021,chen_speeding_2022,blaber_optimal_2023}. 

The rest of the paper is organized as follows.
In Sec.~\ref{sec. Preparation of classical thermal states}, we introduce a transformation between finite- and infinite-temperature Liouvillians for preparing classical thermal states.
In Sec.~\ref{sec. Lindbladian and quantum detailed balance condition}, we extend the transformation to quantum systems to construct Lindbladians for preparing quantum thermal states.
In Sec.~\ref{sec:optimization}, we discuss the optimization of kinetic coefficients to enhance the spectral gap.
In Sec.~\ref{sec. models}, we apply the framework to the kinetic Ising and the transverse-field Ising models, demonstrating enhancements of the spectral gap.
We conclude in Sec.~\ref{sec.conclusion}. 

\section{Preparation of classical thermal states}\label{sec. Preparation of classical thermal states}
Commuting Hamiltonians encompass a broad and computationally challenging class of systems, for example, graph states~\cite{van_den_nest_graph_2008}, toric codes~\cite{alicki_thermalization_2009}, Ising spin-glass~\cite{barahona_computational_1982}. Despite relative simplicity of their Lindbladians, these classical Hamiltonians remain intriguing due to their broad applicability across various physical models. Many of these models are computationally difficult to analyze because of the frustrations in classical Hamiltonians \cite{barahona_computational_1982}. For commuting Hamiltonians, certain bounds on the spectral gap indicate that fast thermalization can be achieved with short range correlations in thermal states~\cite{kastoryano_quantum_2016,bardet_rapid_2023,alicki_thermalization_2009}.

\subsection{Construction of Liouvillians and parent Hamiltonians}
We consider a classical Hamiltonian of the form $\boldsymbol{h}^\mathrm{cl}=\sum_{i=1}^N \boldsymbol{h}^\mathrm{cl}_i$, where all local terms mutually commute and are diagonal in the computational basis $\ket{a}=\ket{a_1,a_2,\ldots,a_N}$, with $a_i$ the local state on site $i$.
The Liouvillian $\boldsymbol{D}$ 
has the maximally mixed state as a left zero eigenvector $\langle \mathbbm{1}|\boldsymbol{D}=0$, where $\ket{\mathbbm{1}}$ denotes the maximally mixed state (unnormalized).
 Under the detailed balance condition, its zero right eigenvector corresponds to the thermal state $\boldsymbol{D}\ket{p_\beta}=0$, with $\ket{p_{\beta}}=e^{-\beta\boldsymbol{h}^{\mathrm{cl}}}/Z(\beta)\ket{\mathbbm{1}}:=\boldsymbol{\rho}_{\beta}\ket{\mathbbm{1}}$ and the partition function $Z(\beta)=\mathrm{tr}(e^{-\beta \boldsymbol{h}^{\mathrm{cl}}})$. 

At infinite-temperature, the thermal state is the maximally mixed state $\ket{\mathbbm{1}}$.
We consider the flip operator {$\tilde{\boldsymbol{A}}_n$} to prepare the maximally mixed state with the tilde indicating infinite temperature. For an eigenstate $\ket{a}$, $\tilde{\boldsymbol{A}}_n\ket{a}$ is a null state or another energy eigenstate of $\boldsymbol{h}^\mathrm{cl}$. Therefore, $\tilde{\boldsymbol{A}}_n$ contains only off-diagonal elements.
The infinite-temperature Liouvillian operator is constructed as 
\begin{align}
\tilde{\boldsymbol{D}}&=\sum_{n}\tilde{\gamma}_{n}(\tilde{\boldsymbol{A}}_{n}-\tilde{\boldsymbol{A}}_{n}^{\top}\tilde{\boldsymbol{A}}_{n}),
\end{align}
where $\tilde{\gamma}_n$ encodes the flip rate, $\top$ denotes the transpose of the flip operator, and the flip operator  $\tilde{\boldsymbol{A}}_n$ contains only $1$ in the off-diagonal elements indicating the transition of the state. 
For classical systems, the probability distribution is real and positive, and both the flip operators and the Liouvillian have real entries.
The second term is the dissipation part, and contains only diagonal elements. 
To ensure that $\ket{\mathbbm{1}}$ is the steady state, i.e. 
$\tilde{\boldsymbol{D}}\ket{\mathbbm{1}}=0$, we require $\tilde{\boldsymbol{D}}=\tilde {\boldsymbol{D}}^\top$, 
which corresponds to the detailed balance condition at infinite temperature 
\begin{align}
\sum_{n}\tilde{\gamma}_{n}\tilde{\boldsymbol{A}}_{n}=\sum_{n}\tilde{\gamma}_{n}\tilde{\boldsymbol{A}}_{n}^{\top}.
\end{align}
This condition is satisfied if the flip operators come in pairs 
$\tilde{\boldsymbol{A}}_{n^\prime}=\tilde{\boldsymbol{A}}_n^\top$ with identical coefficients 
$\tilde{\gamma}_{n^\prime}=\tilde{\gamma}_{n}$.

We now construct the Liouvillian that prepares the classical thermal state $\ket{p_{\beta}}$. The finite-temperature Liouvillian is given by
\begin{align}
\boldsymbol{D}=\sum_{n}\tilde{\gamma}_{n}(\boldsymbol{\rho}_{\beta}^{1/2}\tilde{\boldsymbol{A}}_{n}\boldsymbol{\rho}_{\beta}^{-1/2}-\boldsymbol{A}_{n}^{\top}\boldsymbol{A}_{n}).\label{eq:D_classical}
\end{align}We introduce a finite-temperature flip operator as  
\begin{align}
\boldsymbol{A}_n=\boldsymbol{\rho}_\beta^{1/4}\tilde{\boldsymbol{A}}_n \boldsymbol{\rho}_\beta^{-1/4},\label{eq:flip_op_transform}
\end{align}
and the dissipation part $\boldsymbol{A}_n^\top \boldsymbol{A}_n$ is a diagonal matrix.
The classical detailed balance condition is 
\begin{align}
\boldsymbol{\rho}_{\beta}^{-1}\boldsymbol{D}\boldsymbol{\rho}_{\beta}=\boldsymbol{D}^{\top}.\label{eq:classical detailed balance condition}
\end{align}
Since the dissipation term is diagonal, the classical detailed balance condition only constrains the jump term,
and the dissipation term is diagonal and commute with $\boldsymbol{h}^\mathrm{cl}$.
Equation~\eqref{eq:classical detailed balance condition} leads to $\boldsymbol{D}\boldsymbol{\rho}_{\beta}\ket{\mathbbm{1}}=\boldsymbol{\rho}_{\beta}\boldsymbol{D}^{\top}\ket{\mathbbm{1}}=0
$, ensuring the thermal state as the steady state $\ket{p_\beta}=\boldsymbol{\rho}_{\beta}\ket{\mathbbm{1}}$.

We define the parent Hamiltonian $\boldsymbol{H}=-\boldsymbol{\rho}_{\beta}^{-1/2} \boldsymbol{D}\boldsymbol{\rho}_{\beta}^{1/2}$, which can be written explicitly as
\begin{align}
\boldsymbol{H}=\sum_{n}\tilde{\gamma}_{n}(\boldsymbol{A}_{n}^{\top}\boldsymbol{A}_{n}-\tilde{\boldsymbol{A}}_{n}).
\label{eq:parent_Hamiltonian_classical}
\end{align}
We have used that the dissipative part is diagonal and hence commutes with $\boldsymbol{\rho}_\beta$. The ground state of the parent Hamiltonian is
$\ket{\Psi_g}=\boldsymbol{\rho }_\beta^{1/2}\ket{\mathbbm{1}}.$

For classical many-body systems, the thermal Hamiltonian is a sum of commuting local terms, $\boldsymbol{h}^\mathrm{cl}=\sum_{i=1}^N\boldsymbol{h}_i ^\mathrm{cl}$. We choose local operators $\tilde{\boldsymbol{A}}_{i}$ that induce state changes at site $i$, with $n$ indexing different local transitions. The Liouvillian is constructed as the sum of the local terms $\boldsymbol{D}=\sum_{i=1}^N \tilde{\gamma}_i\boldsymbol{D}_i$, and also the parent Hamiltonian $\boldsymbol{H}=\sum_{i=1}^N \tilde{\gamma}_i\boldsymbol{H}_i $. The kinetic coefficients $\tilde{\gamma}_i$ can be freely chosen while ensuring that the thermal state is the steady state.
Since $\boldsymbol{h}_i^\mathrm{cl}$ commute, the finite-temperature flip operators $\boldsymbol{A}_{i}$ are strictly local.
For example, we can choose a single-spin flip with $\tilde{\boldsymbol{A}}_{i}=\boldsymbol{\sigma}_i^x$. 
 Since $\boldsymbol{h}^\mathrm{cl}$ contains only commute local terms, each dynamical Lie algebra generated by $\tilde{\boldsymbol{A}}_{i}$ remains local, involving only finitely many local operators.

\subsection{Dynamical Lie algebra}

To construct $\boldsymbol{A}_{i}$, we generate the dynamical Lie algebra from $\tilde{\boldsymbol{A}}_{i}$ as 
\begin{align}
\tilde{\boldsymbol{A}}_{i}(t)=e^{i\boldsymbol{h}^{\mathrm{cl}}t}\tilde{\boldsymbol{A}}_{i}e^{-i\boldsymbol{h}^{\mathrm{cl}}t}=\mathfrak{a}_{i}^{\dagger}e^{L_{\mathfrak{a}_i}t}E_{0},
\end{align}
where $\tilde{\boldsymbol{A}}(t)$ represents the Heisenberg evolution of  $\tilde{\boldsymbol{A}}$ under the classical Hamiltonian $\boldsymbol{h}^\mathrm{cl}$ at time $t$, and $\mathfrak{a}_{i}^{\dagger}=(\mathfrak{a}_{i}^{(0)},\mathfrak{a}_{i}^{(1)},\ldots\mathfrak{a}_{i}^{(l_{\mathfrak{a}_i}-1)})$  is a basis of the Krylov space with $\mathfrak{a}_i^{(0)}=\tilde{\boldsymbol{A}}_i$. The inner product is defined as the normalized Hilbert–Schmidt form \begin{align}
    \left\langle\boldsymbol{A},\boldsymbol{B}\right\rangle=\mathrm{tr}(\boldsymbol{A}^\dagger\boldsymbol{B})/d, \label{eq:inner_of_two_operators}
\end{align} with the dimension $d$ of the Hilbert space. Here $E_0=(1,0,0,\ldots,0)^\top$ is a column vector and the operators are Hermitian $\mathfrak{a}_{i}^{(l)\dagger}=\mathfrak{a}_{i}^{(l)}$ and orthonormal $\langle \mathfrak{a}_{i}^{(l)},\mathfrak{a}_{i}^{(k)}\rangle =\delta_{lk}$.
For the case with commuting local terms, the dimension $l_{\mathfrak{a}_i}$ of the dynamical Lie algebra is small, and we explicitly evaluate it in the kinetic Ising model.

The operators in the basis $\mathfrak{a}_i^\dagger $ are Hermitian and form a closed dynamical Lie algebra generated by their commutators with  the classical Hamiltonian $\boldsymbol{h}^\mathrm{cl}$, i.e.,
\begin{align}
i[\boldsymbol{h}^{\mathrm{cl}},\mathfrak{a}_{i}^{(l)}]=\mathfrak{a}_{i}^{(l-1)}L_{\mathfrak{a}_{i}}^{(l-1,l)}+\mathfrak{a}_{i}^{(l+1)}L_{\mathfrak{a}_{i}}^{(l+1,l)},
\end{align}
where the structure constants $L^{(k,l)}_{\mathfrak{a}_i}$ form a skew-symmetric tridiagonal matrix.
Consequently, we represent the finite-temperature flip operator as 
\begin{align}
\boldsymbol{A}_{i}=\boldsymbol{\rho}_{\beta}^{1/4}\tilde{\boldsymbol{A}}_{i}\boldsymbol{\rho}_{\beta}^{-1/4}=\mathfrak{a}_{i}^{\dagger}e^{iL_{\mathfrak{a}_{i}}\beta/4}E_{0}.
\end{align}
 Finally, the parent Hamiltonian is simplified to 
 \begin{align}
\boldsymbol{H}=\sum_{i}\tilde{\gamma}_{i}(\mathfrak{a}_{i}^{\dagger}e^{-i\frac{L_{\mathfrak{a}_{i}}\beta}{4}}E_{00}e^{-i\frac{L_{\mathfrak{a}_{i}}\beta}{4}}\mathfrak{a}_{i}-\mathfrak{a}_{i}^{(0)}),\label{eq:Hforclassicalmanybodysystems}
 \end{align}
where $E_{00}=E_0 E_0^\top $.

We will further apply the above construction for the 1D classical spin models, where the local dynamical Lie algebras have a simple structure with $l_{\mathfrak{a}_i}=3$ because the classical Hamiltonian $\boldsymbol{h}^\mathrm{cl}$ consists of commuting local terms. Under these conditions, our results recover classical kinetic Ising models and enable the construction of classical Liouvillians and parent Hamiltonians for both single- and multiple-spin flips.

\section{Preparation of quantum thermal states}\label{sec. Lindbladian and quantum detailed balance condition}

Under weak-coupling and Markovian conditions, the dynamics of {an open} quantum many-body system are effectively described by the Lindblad master equation, where the evolution of the system is governed by a quantum many-body Lindbladian~\cite{fazio_many-body_2024,landi_nonequilibrium_2022,thompson_field_2023,sa_symmetry_2023,guo_designing_2024}. 
This framework is essential for studying quantum many-body thermodynamics \cite{fazio_many-body_2024,landi_nonequilibrium_2022}. By effectively modeling dissipation and decoherence, the Lindbladian formalism plays a crucial role in describing and optimizing work extraction~\cite{monsel_energetic_2020}, energy transfer~\cite{PhysRevB.99.035421,landi_nonequilibrium_2022}, and error correction~\cite{reiter_dissipative_2017} in quantum systems.
Additionally, it facilitates the design of noise-resilient quantum protocols by effectively modeling noise processes on near-term devices~\cite{preskill_quantum_2018,bharti_noisy_2022}. The Lindblad framework supports the development of more robust and efficient quantum technologies \cite{chen_efficient_2025}.

For an open quantum system, the state is described by the density matrix $\boldsymbol{\rho}$. Its evolution is governed by the Lindblad master equation $\dot{\boldsymbol{\rho}}=\mathcal{L}[\boldsymbol{\rho}]$. We consider the time-local Lindbladian
\begin{align}
\mathcal{L}[\bullet]=-i[\boldsymbol{K},\bullet]+\sum_{n}\gamma_{n}(\boldsymbol{A}_{n}\bullet \boldsymbol{A}_{n}^{\dagger}-\frac{1}{2}\{\boldsymbol{A}_{n}^{\dagger} \boldsymbol{A}_{n},\bullet\}),\label{eq:Lindblad master eq}
\end{align}which is a completely positive trace preserving superoperator \cite{breuer_theory_2002}.
The first term in Eq.~\eqref{eq:Lindblad master eq} is the unitary term with a Hermitian operator $\boldsymbol{K}$, the second term is the jump term with the Lindblad operator $\boldsymbol{A}_n$, and the last term is the dissipation term. 
The kinetic coefficients $\gamma_{n}$ form a Hermitian, positive-semidefinite matrix that characterizes effective dissipative strengths induced by the environment. 
The adjoint Lindbladian follows as
\begin{align}  
    \mathcal{L}^{\sharp}[\bullet]=i[\boldsymbol{K},\bullet]+\sum_{n}\gamma_{n}(\boldsymbol{A}_{n}^{\dagger}\bullet \boldsymbol{A}_{n}-\frac{1}{2}\{\boldsymbol{A}_{n}^{\dagger}\boldsymbol{A}_{n},\bullet\}),\label{eq:adjoint_Lindbladian}
\end{align}
which satisfies $\left\langle \mathcal{L}[\boldsymbol{\rho}],\boldsymbol{B}\right\rangle =\left\langle \boldsymbol{\rho},\mathcal{L}^{\sharp}[\boldsymbol{B}]\right\rangle $. In particular, one verifies that $\mathcal{L}^\sharp [\boldsymbol{I}]=0 $, where  $\boldsymbol{I}$ denotes the  identity operator.

We prepare the thermal state $\boldsymbol{\rho}_\beta=e^{-\beta \boldsymbol{h}}/Z(\beta)$ for a Hamiltonian $\boldsymbol{h}$ at inverse temperature $\beta$, where $Z(\beta)=\mathrm{tr}(e^{-\beta \boldsymbol{h}})$ is the partition function. 
The Lindbladian $\mathcal{L}$ is required to satisfy the standard quantum detailed balance condition~\cite{fagnola_generators_2010,ramezani_quantum_2018}
\begin{align}
\mathcal{L}^{\sharp}[\bullet]-\boldsymbol{\rho}_{\beta}^{-1/2}\mathcal{L}[\boldsymbol{\rho}_{\beta}^{1/2}\bullet\boldsymbol{\rho}_{\beta}^{1/2}]\boldsymbol{\rho}_{\beta}^{-1/2}=2i[\boldsymbol{Q},\bullet],\label{eq:standard_quantum_detailed_balance_condition}
\end{align}with some Hermitian operator $\boldsymbol{Q}$ commuting with $\boldsymbol{\rho}_\beta$, i.e., $[\boldsymbol{Q},\boldsymbol{\rho}_\beta]=0$. With  $\mathcal{L}^\sharp[\boldsymbol{I}]=0$, the detailed balance condition \eqref{eq:standard_quantum_detailed_balance_condition} implies $\mathcal{L}[\boldsymbol{\rho}_\beta]=0$, so that the thermal state is a steady state of the Lindbladian.
Equation~\eqref{eq:standard_quantum_detailed_balance_condition} implies that $\mathcal{L}$ can be decomposed as
$\mathcal{L}[\bullet]=\mathcal{Q}[\bullet]+\mathcal{D}[\bullet]$,
where the coherent (unitary) part is $\mathcal{Q}[\bullet]=-i[\boldsymbol{Q},\bullet]=-\mathcal{Q}^{\sharp}[\bullet]$
and the dissipative part satisfies 
\begin{align}
\mathcal{D}^{\sharp}[\bullet]=\boldsymbol{\rho}_{\beta}^{-1/2}\mathcal{D}[\boldsymbol{\rho}_{\beta}^{1/2}\bullet\boldsymbol{\rho}_{\beta}^{1/2}]\boldsymbol{\rho}_{\beta}^{-1/2}.\label{eq:D_quantum_detailed_balance}
\end{align}
In the following, we focus on purely dissipative generators  $\mathcal{D}[\bullet]$ that satisfy this quantum detailed balance condition.

\subsection{Construction of Lindbladians and parent Hamiltonians}

We now engineer purely dissipative dynamics to prepare the target thermal state $\boldsymbol{\rho}_\beta$. The construction relies on purely dissipative Lindbladians $\mathcal{D}[\bullet]$, built by engineering Lindblad operators $\boldsymbol{A}_n$ and the auxiliary operator $\boldsymbol{K}$, such that the quantum detailed balance condition is satisfied and the thermal state emerges as the unique steady state. The resulting generators are invariant under time-reversal symmetry~\cite{kwon_reversing_2022}. By choosing local Lindblad operators at infinite temperature, we ensure that their finite-temperature counterparts remain quasilocal in the high-temperature (small-$\beta$) regime~\cite{alhambra_quantum_2023}. 

In the weak-coupling limit, Davies' generators give a family of Lindbladians generators with the transitions between energy eigenstates \cite{davies_markovian_1974}. However, they cannot be represented by local operators in quantum many-body systems with non-commuting Hamiltonians. Our framework provides a method for constructing quasilocal Lindbladians to prepare quantum thermal states, and it reproduces Davies' generators for classical commuting Hamiltonians.

Compared to Refs.~\cite{chen_efficient_2025,gilyen_quantum_2024}, we provide a general construction of the Lindbladians to satisfy the quantum detailed balanced condition: instead of the operator Fourier transform, we require a similar transform by Eq.~\eqref{eq:flip_op_transform}
on the Lindblad operators to construct Lindbladians at finite temperature. 
We note that the locality of the Lindbladians can be improved by applying a filter to the infinite-temperature Lindblad operators $\tilde{\boldsymbol{A}}_n$ \cite{chen_efficient_2025,gilyen_quantum_2024} (see also Appendix~\ref{sec:Remarks on CKG algorithm}).
We show the construction in the following.

At infinite temperature $\beta=0$, the thermal state is the maximally mixed density matrix $\boldsymbol{\rho}_0 \propto \boldsymbol{I}$, and Eq.~\eqref{eq:D_quantum_detailed_balance} simplifies to $\tilde{\mathcal{D}}^\sharp[\bullet]=\tilde{\mathcal{D}}[\bullet]$. 
The tilde denotes the superoperators (or operators) at infinite temperature. 
The infinite-temperature Lindbladian is generally constructed as 
\begin{align}
\tilde{\mathcal{D}}[\bullet]=\sum_{n}\tilde{\gamma}_{n}(\tilde{\boldsymbol{A}}_{n}\bullet\tilde{\boldsymbol{A}}_{n}^{\dagger}-\frac{1}{2}\{\tilde{\boldsymbol{A}}_{n}^{\dagger}\tilde{\boldsymbol{A}}_{n},\bullet\}).
\end{align}
The coefficients $\tilde{\gamma}_{n}$ and Lindblad operators $\tilde{\boldsymbol{A}}_n$ are chosen to satisfy $\tilde{\mathcal{D}}^{\sharp}[\bullet]=\tilde{\mathcal{D}}[\bullet]$, i.e., 
\begin{align}
\sum_{n}\tilde{\gamma}_{n}\tilde{\boldsymbol{A}}_{n}\bullet\tilde{\boldsymbol{A}}_{n}^{\dagger}=\sum_{n}\tilde{\gamma}_{n}\tilde{\boldsymbol{A}}_{n}^{\dagger}\bullet\tilde{\boldsymbol{A}}_{n}. \label{eq:infinitetemperature_jump_detailed_balance}
\end{align} 
For example, it suffices to take pairs of Lindblad operators  $\tilde{\boldsymbol{A}}_{n^\prime}=\tilde{\boldsymbol{A}}_n^\dagger$ with identical coefficients $\tilde{\gamma}_{n^\prime}=\tilde{\gamma}_{n}$ for each pair of jumps.

To prepare the thermal state $\boldsymbol{\rho}_\beta$, the finite-temperature Lindbladian is constructed as 
\begin{align}
\mathcal{D}[\bullet]&=-i[\boldsymbol{K},\bullet]-\frac{1}{2}\{\boldsymbol{R},\bullet\}+\sum_{n}\tilde{\gamma}_{n}\boldsymbol{A}_{n}\bullet\boldsymbol{A}_{n}^{\dagger},\label{eq:Dissipator_detailed_balance_condition}
\end{align}
where for convenience we introduce the dissipation operator
\begin{align}
    \boldsymbol{R}=\sum_n \tilde{\gamma}_n \boldsymbol{A}_n ^\dagger \boldsymbol{A}_n.
\end{align}
The Lindblad operator $\boldsymbol{A}_n$ transforms as  
\begin{align}
\boldsymbol{A}_n=\boldsymbol{\rho}_\beta^{1/4}\tilde{\boldsymbol{A}}_n \boldsymbol{\rho}_\beta ^{-1/4}=\tilde{\boldsymbol{A}}_{n}(\frac{i\beta}{4}),\label{eq:A_k_1}
\end{align}
where the operator is propagated forward by an imaginary-time shift of $i\beta/4$.

As shown in Ref.~\cite{chen_efficient_2025}, the coherent part $\boldsymbol{K}$ can be chosen such that the Lindbladian $\mathcal{D}[\bullet]$ satisfies the quantum detailed balance condition \eqref{eq:D_quantum_detailed_balance}, provided that the jump operators obey this condition.
Using Eq.~\eqref{eq:D_quantum_detailed_balance}, the coherent part $\boldsymbol{K}$ is determined by 
\begin{align}
\{\boldsymbol{\rho}_{\beta}^{-\frac{1}{2}},\boldsymbol{\rho}_{\beta}^{\frac{1}{4}}\boldsymbol{K}\boldsymbol{\rho}_{\beta}^{\frac{1}{4}}\}=\frac{i}{2}[\boldsymbol{\rho}_{\beta}^{-\frac{1}{2}},\boldsymbol{\rho}_{\beta}^{\frac{1}{4}}\boldsymbol{R}\boldsymbol{\rho}_{\beta}^{\frac{1}{4}}],
\label{eq:equation_to_solve_K}
\end{align}
where the derivation is given in Appendix~\ref{apppendix D:derivation to G}.
Therefore, Eqs.~\eqref{eq:Dissipator_detailed_balance_condition}-\eqref{eq:equation_to_solve_K} define a general transformation that maps infinite-temperature Lindbladians $\mathcal{D}$ to their finite-temperature counterparts $\tilde{\mathcal{D}}$.

The parent Hamiltonian can be defined for the dissipative Lindbladian as 
\begin{align}
    \mathcal{H}[\bullet]&:=-\boldsymbol{\rho}_{\beta}^{-1/4}\mathcal{D}[\boldsymbol{\rho}_{\beta}^{1/4}\bullet\boldsymbol{\rho}_{\beta}^{1/4}]\boldsymbol{\rho}_{\beta}^{-1/4},
    \label{eq:parent_H_quantum}
\end{align}
and the detailed balance condition~\eqref{eq:D_quantum_detailed_balance} leads to
\begin{align}
    \mathcal{H}^{\sharp}[\bullet]=\mathcal{H}[\bullet].
\end{align}
The total Lindbladian satisfies the quantum detailed balance condition, ensuring frustration-freeness of the parent Hamiltonian.

Using the construction of the Lindbladian by Eq. ~\eqref{eq:Dissipator_detailed_balance_condition}, we construct the parent Hamiltonian in the thermofield-double space as 
\begin{align}
\mathcal{H}[\bullet]=\frac{1}{2}\left\{ \boldsymbol{N},\bullet\right\} -\sum_{n}\tilde{\gamma}_{n}\tilde{\boldsymbol{A}}_{n}\bullet\tilde{\boldsymbol{A}}_{n}^{\dagger},\label{eq:parent_H_thermofielddouble}
\end{align}
where $\boldsymbol{N}$ is constructed to ensure $\boldsymbol{\rho}_\beta^{1/2}$ is the ground state of the parent Hamiltonian
\begin{align}
\mathcal{H}[\boldsymbol{\rho}_{\beta}^{1/2}]=\frac{1}{2}\{\boldsymbol{N},\boldsymbol{\rho}_{\beta}^{1/2}\}-\sum_{n}\tilde{\gamma}_{n}\tilde{\boldsymbol{A}}_{n}\boldsymbol{\rho}_{\beta}^{1/2}\tilde{\boldsymbol{A}}_{n}^{\dagger}=0,
\end{align}
i.e.,\begin{align}
\boldsymbol{\rho}_{\beta}^{-1/4}\boldsymbol{N}\boldsymbol{\rho}_{\beta}^{1/4}+\boldsymbol{\rho}_{\beta}^{1/4}\boldsymbol{N}\boldsymbol{\rho}_{\beta}^{-1/4}=2\boldsymbol{R}.\label{eq:equation_to_solve_N}
\end{align}
For small systems, $\boldsymbol{K}$ in ${\mathcal{D}}[\bullet]$ and $\boldsymbol{N}$ in $\mathcal{H}[\bullet]$ can be obtained by solving the linear equations \eqref{eq:equation_to_solve_K}
and \eqref{eq:equation_to_solve_N}, respectively. As the system size increases, Eqs.~\eqref{eq:equation_to_solve_K} and~\eqref{eq:equation_to_solve_N} become intractable to solve directly. We therefore exploit the dynamical Lie algebra generated by $\boldsymbol{R}$ to construct $\boldsymbol{K}$ and $\boldsymbol{N}$.

\subsection{Time domain representation} 

In the energy domain, the operators can be decomposed according to energy change
\begin{align}
\boldsymbol{K}=\sum_{\nu}\sum_{\nu=E_{b}-E_{a}}\ket{b}\bra{b}\boldsymbol{K}\ket{a}\bra{a}:=\sum_{\nu}\boldsymbol{K}_{\nu}. 
\end{align}
where $\nu$ denotes the energy change of two energy eigenstates with respect to the Hamiltonian $\boldsymbol{h}$.
The coherent part is solved in the energy domain as 
\begin{align}
    \boldsymbol{K}=\sum_{\nu}\frac{i}{2}\tanh(\frac{\beta\nu}{4})\boldsymbol{R}_{\nu}.\label{eq:Kenergydomain}
\end{align}
We introduce the Fourier transform as 
\begin{align}
    \boldsymbol{R}(t)&=\sum_\nu e^{i\nu t}\boldsymbol{R}_\nu, \label{eq:Rt}\\
    \boldsymbol{R}_\nu&=\frac{1}{2\pi}\int_{-\infty}^\infty e^{-i\nu t}\boldsymbol{R}(t)dt. \label{eq:Rnu}
\end{align}
Substituting Eq.~\eqref{eq:Rnu} into Eq.~\eqref{eq:Kenergydomain}, the time-domain representation is obtained as \cite{chen_efficient_2025}
\begin{align}
\boldsymbol{K}=\mathrm{p.v.}\int_{-\infty}^{\infty}\;\frac{\boldsymbol{R}(t)}{\beta\sinh(2\pi t/\beta)}dt,\label{eq:Ktimedomain}
\end{align}
where $\mathrm{p.v.}$ denotes the principal value of the integral over $t$.
The time domain representation enables the construction of  $\boldsymbol{K}$ without diagonalizing $\boldsymbol{R}$.

To construct the parent Hamiltonian $\mathcal{H}[\bullet]$, we represent $\boldsymbol{N}$ in the energy domain and obtain 
\begin{align}
\boldsymbol{N}=\sum_{\nu}\frac{\boldsymbol{R}_{\nu}}{\cosh(\beta\nu/4)}.
\end{align}
Its time-domain representation can be written as
\begin{align}
\begin{aligned}
\boldsymbol{N}&=\int_{-\infty}^{\infty}\frac{2\boldsymbol{R}(t)}{\beta\cosh(2\pi t/\beta)}dt.\label{eq:Ntermquantum}
\end{aligned}
\end{align}
We also remark that Eq.~\eqref{eq:Ntermquantum} can be used to evaluate the symmetric logarithmic derivative, which is useful for studying quantum metrology with tensor networks \cite{chabuda_tensor-network_2020}.

For quantum many-body systems, we construct $\boldsymbol{A}_n$ and $\boldsymbol{K}$ as  quasilocal operators by exploiting the dynamical Lie algebra generated from $\tilde{\boldsymbol{A}}_n$ and $\boldsymbol{R}$. Moroever, by selecting $\tilde{\boldsymbol{A}}_n$ as the transition between different energy eigenstates, the Lindbladian recovers Davies' generators \cite{davies_markovian_1974}, but such Lindblad operators will be nonlocal operators for quantum noncommuting Hamiltonians.

\subsection{Dynamical Lie algebra}

In quantum many-body systems, the Hamiltonian consists of local terms $\boldsymbol{h} = \sum_{i=1}^N \boldsymbol{h}_i$, each supported on neighboring sites within a finite range. 
We construct $\mathcal{D}_i[\bullet]$ from local operators $\tilde{\boldsymbol{A}}_{i}$, and combine them as
$\mathcal{D} [\bullet]= \sum_{i=1}^N  \tilde{\gamma}_i \mathcal{D}_i[\bullet]$, where $\tilde{\gamma}_i$ are kinetic coefficients.
At finite temperature, the dissipative Lindbladian $\mathcal{D} [\bullet]$ becomes quasilocal, but can still be built from the dynamical Lie algebra generated by the Lindblad operators $\tilde{\boldsymbol{A}}_{i}$ at infinite temperature. 

We choose $\tilde{\boldsymbol{A}}_i$ as a local operator, for example, local Pauli operators in spin models. 
Given the Hamiltonian $\boldsymbol{h}$, the corresponding dynamical Lie algebra is generated from $\tilde{\boldsymbol{A}}_i$. In the Heisenberg picture, the time evolution of the operators is given by
\begin{align}
    \tilde{\boldsymbol{A}}_i (t)=e^{i\boldsymbol{h}t}\tilde{\boldsymbol{A}_i}e^{-i\boldsymbol{h}t}=\mathfrak{a}_{i}^{\dagger} e^{L_{\mathfrak{a}_i} t} E_0,\label{eq:Atildetquantum}
\end{align}
where $E_0 =(1,0,0,\ldots)^\top$ is a column vector. 
The finite-temperature Lindblad operator can be constructed as 
\begin{align}
\boldsymbol{A}_{i}&=\tilde{\boldsymbol{A}}_{i}(\frac{i\beta}{4})=\mathfrak{a}_{i}^{\dagger}e^{i\beta L_{\mathfrak{a}_{i}}/4}E_{0}=E_{0}^{\top}e^{-i\beta L_{\mathfrak{a}_{i}}/4}\mathfrak{a}_{i},\label{eq:A_{i}dynamicalLie}\\\boldsymbol{A}_{i}^{\dagger}&=\tilde{\boldsymbol{A}}_{i}(-\frac{i\beta}{4})=\mathfrak{a}_{i}^{\dagger}e^{-i\beta L_{\mathfrak{a}_{i}}/4}E_{0}.\label{eq:A_idagger_dynamicalLie}
\end{align}
We denote the basis by $\mathfrak{a}_{i}^{\dagger}=(\mathfrak{a}_{i}^{(0)},\mathfrak{a}_{i}^{(1)},\ldots\mathfrak{a}_{i}^{(l_{\mathfrak{a}_{i}}-1)})$, written as a row vector. Since the basis of the dynamical Lie algebra is Hermitian, each element satisfies $\mathfrak{a}^{(l)\dagger}_i = \mathfrak{a}^{(l)}_i$. Then we can write the dissipation term as 
\begin{align}
\boldsymbol{R}=\sum_{i=1}^N\tilde{\gamma}_{i}\boldsymbol{A}_{i}^{\dagger}\boldsymbol{A}_{i}=\sum_{i=1}^N\tilde{\gamma}_{i}\mathfrak{a}_{i}^{\dagger}e^{-i\beta L_{\mathfrak{a}_i}/4}E_{00}e^{-i\beta L_{\mathfrak{a}_i}/4}\mathfrak{a}_{i},
\end{align}
where $E_{00}=E_0E_0^{\top}$. We solve $L_{\mathfrak{a}_i}$ with the Krylov-space expansion truncated at $l_{\mathfrak{a}_i}$ (see Appendix~\ref{sec:Krylov space expansion for operator dynamics}) as the dynamical Lie algebra for quantum systems can be of large dimension. This allows us to evaluate the matrix $e^{-i\beta L_{\mathfrak{a}_i}/4}E_{00}e^{-i\beta L_{\mathfrak{a}_i}/4}$. The adequacy of the truncation can be verified by checking that the neglected matrix elements are sufficiently small.

To construct Lindbladians and parent Hamiltonians for quantum thermal states, we solve the dynamical Lie algebra generated by the dissipation operator $\boldsymbol{R}$ to ensure the quantum detailed balance condition. We first rewrite 
the dissipation part as
\begin{align}
\boldsymbol{R}=r\boldsymbol{I}+r_{0}\hat{\boldsymbol{R}}, 
\end{align}
where $\hat{\boldsymbol{R}}$ has the norm 1. When the Hamiltonian and kinetic coefficients are translationally invariant, the dynamical Lie algebra $\mathfrak{r}^\dagger$ inherits the translational symmetry.
We construct the dynamical Lie algebra from $\hat {\boldsymbol{R}}$ to represent 
\begin{align}
\hat{\boldsymbol{R}}(t)&=e^{i\boldsymbol{h}t}\hat{\boldsymbol{R}}e^{-i\boldsymbol{h}t}=\mathfrak{r}^{\dagger}e^{L_{\mathfrak{r}} t}E_0,
\end{align}
and we obtain 
\begin{align}
\boldsymbol{R}(t)=r\boldsymbol{I}+r_{0}\mathfrak{r}^{\dagger}e^{L_{\mathfrak{r}} t}E_0.
\end{align}
Equations~\eqref{eq:Ktimedomain} and~\eqref{eq:Ntermquantum} are represented with the dynamical Lie algebra into 
\begin{align}
    \boldsymbol{K}=
    r_{0}\mathfrak{r}^{\dagger}\frac{\beta}{2}\tan(\frac{\beta L_{\mathfrak{r}}}{4})E_{0},\label{eq:Ktimedomainnew}
\end{align}
and 
\begin{align}
\boldsymbol{N}&=r+r_{0}\mathfrak{r}^{\dagger}\sec(\frac{\beta L_{\mathfrak{r}}}{4})E_0,\label{eq:Ntermliealgebra}
\end{align}
which can be constructed into a matrix product operator after truncating the expansion at $l_\mathfrak{r}$.  

\subsection{Tensor-network construction of parent Hamiltonians}

We reshape the parent Hamiltonian into an operator acting on the thermofield-double space (see Appendix~\ref{Appedix_A_superoperator}), whose ground state corresponds to the thermofield-double state.
\begin{align}
\ket{\boldsymbol{\rho}_{\beta}^{1/2}}=\frac{e^{-\frac{\beta}{2}\boldsymbol{h}}}{\sqrt{Z(\beta)}}\ket{\Phi},\label{eq:purified_state}
\end{align}
where $\ket{\Phi}=\sum_a\ket{a}_L\otimes\ket{a}_R$ is the (unnormalized) maximally entangled state between the physical and auxiliary systems, and $\ket{a}$ is the eigenstate of $\boldsymbol{h}$.
The purified state (Eq.~\eqref{eq:purified_state}) encodes the thermal state $\boldsymbol{\rho}_\beta$, and the spectrum of $\mathcal{H}$ corresponds to that of the Lindbladian $\mathcal{D}$ up to a sign, reflecting its relaxation rates.

We construct the parent Hamiltonian explicitly as
\begin{align}
\mathcal{H}=\frac{1}{2}(\boldsymbol{N}\otimes \boldsymbol{I}+\boldsymbol{I}\otimes \boldsymbol{N}^{\top})-\sum_{i=1}^{N}\tilde{\gamma}_{i}\tilde{\boldsymbol{A}}_{i}\otimes\tilde{\boldsymbol{A}}_{i}^{*},
\end{align}
where a transpose acts on the operators in the auxiliary space.
In one-dimensional quantum spin models, $\mathcal{H}$ takes the form of a spin-ladder Hamiltonian. Using the basis of the dynamical Lie algebra, the parent Hamiltonian can be efficiently represented as a matrix product operator, allowing computing its spectral gap via DMRG. We further optimize the kinetic coefficients $\tilde{\gamma}_i$ to enhance the spectral gap of Lindbladians.

\section{Spectral gap optimization}\label{sec:optimization}

One of the central questions in nonequilibrium statistical mechanics is how fast a system reaches thermal equilibrium and, if possible, how to accelerate this relaxation process.
The minimum relaxation rate is determined by the spectral gap of the Lindbladian. 
However, calculating this gap exactly is infeasible for generic many-body systems. For instance, in a system of $n$ qubits, this requires finding the eigenvalue of the Lindbladian superoperator with the smallest nonzero real part in absolute value. The superoperator is represented by a $d^2 \times d^2$-matrix, where $d=2^n$.
This task is impossible beyond a few qubits, therefore, it is crucial to identify more scalable methods. 

In the previous sections, we have constructed Liouvillians and Lindbladians for preparing both classical and quantum thermal states. The kinetic coefficients $\tilde{\gamma}_n$ contain free parameters that do not alter the steady-state Gibbs distribution, and can be optimized to enhance the spectral gap and thereby boost thermalization. We compare the optimized coefficients with the canonical uniform choice. To remove any trivial scaling, we introduce a normalized gap to formulate the corresponding optimization problem in the following.

We consider a system with Hilbert-space dimension $d$: in the quantum case the state is represented by a $d \times d$ density matrix $\boldsymbol{\rho}$, while in the classical case it is a $d$-dimensional probability vector $\ket{p}$. 
Here we first focus on the quantum case.
To boost thermalization in preparing thermal states, we optimize the dissipative generator $\mathcal{D}$ to enhance its spectral gap.  
Under the quantum detailed-balance condition, 
the spectrum of $\mathcal{D}$ is real and non-positive. We therefore write the eigenvalues as $-\lambda_j$ with $\lambda_j \ge 0$, where $\lambda_j$ denote the corresponding relaxation rates.
The zero mode $\lambda_0 = 0$ corresponds to the steady state (the ground state of $\mathcal{H}$). The eigenvalues are ordered increasingly as $\lambda_0,\lambda_1,\ldots,\lambda _{d^2-1}$, and the thermal state $\boldsymbol{\rho}_\beta$ corresponds to the non-degenerate zero mode.
The spectral gap is then the smallest nonzero eigenvalue of $-\mathcal{D}$ or $\mathcal{H}$, i.e., $\lambda_1$.
Here we focus on the case where the system has a unique steady state, but the framework can be in principle generalized to degenerate steady states. 

For a given choice of kinetic coefficients $\tilde{\gamma}_n$, we can formulate the SDP that certifies a lower bound on $\lambda_1$ as \cite{rai_hierarchy_2024,cruz_preparation_2022}
\begin{align}
\begin{split}
\max_x &\quad x, 
\\\mathrm{subject\;to}&\quad\mathcal{H}^{2}-x \mathcal{H}\succeq0.
\end{split}
\end{align}
However, since $\mathcal{H}$ depends linearly on the kinetic coefficients $\tilde{\gamma}_n$, the subsequent optimization over $\tilde{\gamma}_{n}$ becomes nonlinear.
The optimization problem for the spectral gap can thus be formulated as
\begin{align}
\begin{split}
\max_{\tilde{\gamma}_{n}}&\quad\Delta=\frac{\lambda_{1}}{\tau(\mathcal{H})}=\frac{\lambda_{1}}{\sum_{j=0}^{d^{2}-1}\lambda_{j}/d^{2}},\\
\mathrm{subject\;to}&\quad\sum_{n}\tilde{\gamma}_{n}\tilde{\boldsymbol{A}}_{n}\bullet\tilde{\boldsymbol{A}}_{n}^{\dagger}=\sum_{n}\tilde{\gamma}_{n}\tilde{\boldsymbol{A}}_{n}^{\dagger}\bullet\tilde{\boldsymbol{A}}_{n},
\end{split}\label{eq:quantumoptimizationproblem}
\end{align}
where $\Delta$ is the normalized gap, $\tau(\mathcal{H})=\sum_{j=0}^{d^2-1}\lambda_j/d^2 $ denotes the normalized trace of the superoperator, and $\mathcal{H}$ is constructed according to Eq.~\eqref{eq:parent_H_thermofielddouble}. The constraint in Eq.~\eqref{eq:quantumoptimizationproblem} corresponds to the detailed balance condition, which can be satisfied, for example, by choosing identical kinetic coefficients $\tilde{\gamma}_{n}=\tilde{\gamma}_{n^\prime}$ for each pair of Lindblad operators at infinite temperature.

Additional constraints can be introduced, since for large systems arbitrary Lindblad operators are not experimentally accessible. 
In such cases, the admissible operators are restricted to specific forms or are constrained by the locality inherent in many-body systems. When these constraints are present, optimizing the spectral gap becomes nontrivial. For example, the total Lindbladian can be assembled from local contributions, each associated with a set of kinetic coefficients $\tilde{\gamma}_i$. The spectral gap can be further enhanced by tuning the coefficients $\tilde{\gamma}_i$, which control the weights of the local terms.
When exact calculation of the spectral gap is infeasible for many-body systems, combining a variational upper bound with a certified lower bound narrows the interval containing the true spectral gap. 

The gap-optimization technique can similarly be adapted to classical Liouvillians. 
The eigenvalues are ordered increasingly as $\lambda_0,\lambda_1,\cdots,\lambda_{d-1}$. 
Here the normalization constant $d$ differs because we shift from the thermofield-double space to the probability-distribution space.
The optimization of the normalized gap $\Delta$ is formulated as
\begin{align}
\begin{split}
\max_{\tilde{\gamma}_{n}}\quad&\Delta=\frac{\lambda_{1}}{\tau(\boldsymbol{H})}=\frac{\lambda_{1}}{\sum_{j=0}^{d-1}\lambda_{j}/d},\\\mathrm{subject\;to}&\quad\sum_{n}\tilde{\gamma}_{n}\tilde{\boldsymbol{A}}_{n}=\sum_{n}\tilde{\gamma}_{n}\tilde{\boldsymbol{A}}_{n}^{\top},
\end{split}
\end{align}
where $\Delta$ is the normalized gap and $\boldsymbol{H}$ is constructed according to Eq.~\eqref{eq:parent_Hamiltonian_classical}. The normalized trace of the operator is $\tau(\boldsymbol{H})=\mathrm{tr}(\boldsymbol{H})/d=\sum_{j=0}^{d-1}\lambda_{j}/d$. 
Both the spectral gap and $\tau(\boldsymbol{H})$ are piecewise-smooth functions of the kinetic coefficients, allowing gradient-based optimization.
In this paper, we focus on finite-dimensional systems, where the normalized gap is well defined since all operators are trace-class. For infinite-dimensional systems, the normalized trace $\tau(\boldsymbol{H})=\mathrm{tr}(\boldsymbol{H})/\mathrm{tr}(\boldsymbol{I})$ is no longer available because $\mathrm{tr}(\boldsymbol{I})=\infty$, but our method of optimizing the spectral gap $\lambda_1$ may still be  applicable.

\subsection{Symmetries}\label{subsection:symmetries}
In many-body systems, determining the kinetic coefficients $\tilde{\gamma}_{i}$ of a Lindbladian often involves numerous free parameters. Fortunately, the symmetries of the thermal state can help reduce these parameters, simplifying the optimization of Lindbladians. To enhance the spectral gap $\Delta$, the symmetry group of the Lindbladian $\mathcal{D}$ must exactly match that of the thermal state $\boldsymbol{\rho}_\beta$. It can be neither larger nor smaller.

Suppose we have a thermal state with a symmetry $G$, represented by the unitary $\boldsymbol{U}_g$ acting on the Hilbert space for each $g\in G$. 
This symmetry $G$ is general; for example, it can include translational symmetry or discrete symmetries such as the flip symmetry in kinetic Ising models.
The thermal state remains unchanged under this symmetry, satisfying $\boldsymbol{U}_g \boldsymbol{\rho}_\beta \boldsymbol{U}_g^\dagger=\boldsymbol{\rho}_\beta$. 

To impose this symmetry on the Lindbladian superoperator, we transform it as 
\begin{align}
\mathcal{D}_{g}[\bullet]=\boldsymbol{U}_{g}^{\dagger}\mathcal{D}[\boldsymbol{U}_{g}\bullet\boldsymbol{U}_{g}^{\dagger}]\boldsymbol{U}_{g},\label{eq:symmetry action in L}
\end{align}
where $\boldsymbol{\rho}_\beta$ remains the steady state, and all the operators appeared in the Lindbladian is transformed by $\boldsymbol{U}_g$. 
Importantly, this unitary transformation does not alter the normalized gap of $\mathcal{D}$. 
We construct a symmetric Lindbladian as 
\begin{align}
\mathcal{D}_{\mathrm{sym}}[\bullet]=\frac{\int_{G}d\mu_{g}\boldsymbol{U}_{g}^{\dagger}\mathcal{D}[\boldsymbol{U}_{g}\bullet\boldsymbol{U}_{g}^{\dagger}]\boldsymbol{U}_{g}}{\int_{G}d\mu_{g}1},\label{eq:symmetric Lindbladian}
\end{align}where $\mu_g$ is the Haar measure of the group $G$. If $G$ is a finite group, Eq.~\eqref{eq:symmetric Lindbladian} is simplified to the average over all group elements.
According to Weyl's inequality~\cite{weyl_asymptotische_1912}, the spectral gap is a concave function of the underlying parent Hamiltonians~\cite{rai_spectral_2025}. This concavity also extends to Lindbladians, implying that the normalized gap of the symmetric Lindbladian satisfies
\begin{align}
    \Delta(\mathcal{D}_\mathrm{sym})\geq \Delta(\mathcal{D}).
\end{align}
Therefore, to enhance the spectral gap, the Lindbladian $\mathcal{D}$ should preserve the same symmetries as the thermal state $\rho_\beta$. 
Under the transformation in Eq.~\eqref{eq:symmetry action in L}, $\mathcal{D}_g$ modifies the kinetic coefficients $\tilde{\gamma}_{n}$. The symmetry group $G$ thus acts on these coefficients. The optimal kinetic coefficients $\tilde{\gamma}^{\mathrm{opt}}_{n}$  remain invariant under this group action.

When the symmetry group of the Lindbladian contains that of the thermal state as a proper subgroup, additional steady states can be generated from $\boldsymbol{U} \boldsymbol{\rho}_\beta \boldsymbol{U}^\dagger$, where $\boldsymbol{U}$ is a unitary representation of the symmetry group of the Lindbladian. The spectral gap is thus closed due to degenerate steady states, and the spontaneously symmetry breaking occurs~\cite{minganti_spectral_2018}. 

In particular, we consider the Hamiltonian that generates the symmetry group
\begin{align}
    \boldsymbol{U}_t=e^{-i \boldsymbol{h}t}.
\end{align}
The thermal state is invariant under this action $\boldsymbol{\rho}_\beta=\boldsymbol{U}_t \boldsymbol{\rho}_\beta \boldsymbol{U}_t^\dagger$. In finite-dimensional systems, this one-parameter unitary group is compact and therefore admits a well-defined Haar measure. 
We therefore expect that an optimal Lindbladian also takes the symmetrized form
\begin{align}
\mathcal{D}_{\mathrm{sym}}[\bullet]=\frac{1}{T}\int_{0}^{T}dt\boldsymbol{U}_{t}^{\dagger}\mathcal{D}[\boldsymbol{U}_{t}\bullet\boldsymbol{U}_{t}^{\dagger}]\boldsymbol{U}_{t},
\end{align}
where $T$ is the fundamental period of the unitary group $\boldsymbol{U}_T=\boldsymbol{U}_0$. However, for many-body systems, the fundamental period $T$ typically scales exponentially with the system size $N$, and for large $t$ the unitary evolution $\boldsymbol{A}(t)=e^{i\boldsymbol{h}t}\boldsymbol{A}e^{-i\boldsymbol{h}t}$ leads to operator growth and thus destroys locality.
Averaging the time-evolved dissipative generator $\mathcal{D}_t[\bullet] 
:= \boldsymbol{U}_{t}^{\dagger}\mathcal{D}[\boldsymbol{U}_{t}\bullet\boldsymbol{U}_{t}^{\dagger}]\boldsymbol{U}_{t}$ over $\boldsymbol{U}_{t}$ associates independent Lindblad operators with distinct energy changes and ensures separate Lindblad operators for decoherence and dissipation processes.
Therefore, if locality is not taken into account, the optimal Lindbladian is in the form of Davies' generators \cite{davies_markovian_1974}: transitions involving different energy changes are driven by independent Lindblad operators.

In practice, we may impose locality constraints on the operators appearing in the Lindbladian.
The symmetries considered here include the flip symmetry and the translational symmetry. 
With only a limited set of Lindblad operators under such constraints, the Davies' generators cannot be recovered after optimization in general.

\subsection{Optimizing gap for single-body systems}

We now formulate the optimization of a Lindbladian in a single-body system to maximize the normalized gap $\Delta$ for given total dissipation strength, without imposing the constraint of locality. 

For a $d$-dimension single-body system, when arbitrary Lindblad operators are accessible, we find the optimal gap is 
\begin{align}
\Delta^\mathrm{opt}=\frac{d^2 }{d^2 -1}, \label{eq:r_opt_quantum}
\end{align}
where the spectrum contains a single zero eigenvalue for the thermal state, and the rest $d^2-1$ eigenvalues are identical to $\Delta^\mathrm{opt}$.

For a classical $d$-state system, the optimal gap becomes
\begin{align}
\Delta^\mathrm{opt}=\frac{d}{d-1}. \label{eq:r_opt_classical}
\end{align}
The different coefficient comes from the dimensions of Hilbert spaces. These bounds provide the maximum upper limit that the spectral gap can reach in the absence of locality and other constraints on Lindblad operators.

\subsection{Optimizing gap for many-body systems}
To prepare thermal states $\boldsymbol{\rho}_\beta$ for a many-body systems, we choose $\tilde{\boldsymbol{A}}_i$ as local operators, such as Pauli operators on one or two sites of a spin model. Locality can be further improved by filtering in the energy domain via the operator Fourier transform \cite{chen_efficient_2025} (see Appendix \ref{sec:Remarks on CKG algorithm}). 
Under these locality constraints, the previous bounds~\eqref{eq:r_opt_quantum} and \eqref{eq:r_opt_classical} cannot be achieved.

We optimize the kinetic coefficients $\tilde{\gamma}_{i}$ to maximize the normalized gap $\Delta$. 
As discussed in Sec.~\ref{subsection:symmetries}, global and discrete symmetries can be employed to encode the kinetic coefficients $\tilde{\gamma}_{i}$, thereby reducing the number of free parameters. The gap is evaluated using the parent  Hamiltonian in the thermofield-double space, $\lambda_1=\bra{\Psi_1}\mathcal{H} \ket{\Psi_1}$, and its derivatives with respect to the kinetic coefficients follow from the Hellmann–Feynman theorem as 
\begin{align}
\frac{\partial\lambda_{1}}{\partial\tilde{\gamma}}=&\bra{\Psi_{1}}\frac{\partial\mathcal{H}}{\partial\tilde{\gamma}}\ket{\Psi_{1}},
\end{align}
which also applies to the normalized gap when the trace of the parent Hamiltonian is fixed.
The first excited state $\ket{\Psi_{1}}$ can be computed exactly for small systems and variationally with the DMRG for larger systems. When exact diagonalization is not feasible, the spectral gap can still be certified from below using SDP~\cite{cruz_preparation_2022,rai_hierarchy_2024, kull_lower_2024,robichon_bootstrapping_2024,mortimer_certifying_2025}. As a demonstration, we employ SDP to certify the spectral gap of Liouvillians in the single-spin flip kinetic Ising model.

\section{Models}\label{sec. models}
We apply our method to prepare thermal states and optimize the normalized gap for both classical and quantum models.
For classical systems, we construct the Liouvillian and optimize the normalized gap for the kinetic Ising models with both single-spin and double-spin flips. 
For quantum systems, we construct the Lindbladian and optimize the normalized gap in the 1D transverse-field Ising model with single-site Pauli Lindblad operators.

\subsection{1D single-spin flip kinetic Ising model}\label{sec:1D_kinetic_Ising}

We consider the classical kinetic Ising model \cite{glauber_time-dependent_1963,haake_universality_1980}, which describes the relaxation dynamics of spins on a lattice. 
Each spin exists in one of two states (up or down, i.e. $\boldsymbol{\sigma}_i^z=\pm1$), and flips between these states over time, driving the system toward the thermal state. The model evolves based on local interactions, and the flip rates depend on the state of its neighboring spins, and they are chosen to satisfy detailed balance. Also, there are free parameters in the kinetic coefficients, leaving space for spectral gap optimization. 
The classical Hamiltonian consists of commuting local terms.

In the following, we focus on optimizing the spectral gap of the Liouvillian for the 1D kinetic Ising model with single-spin flip.  The classical Hamiltonian of the 1D Ising model is given by
\begin{align}
\boldsymbol{h}^{\mathrm{cl}}=-\sum_{i=1}^{N}J\boldsymbol{\sigma}_{i}^{z}\boldsymbol{\sigma}_{i+1}^{z}.\label{eq:1D Ising Hamiltonian}
\end{align}
We consider the periodic boundary condition, setting $\boldsymbol{\sigma}_{N+1}^z=\boldsymbol{\sigma}_{1}^z$ in the last local term. The Hamiltonian~\eqref{eq:1D Ising Hamiltonian} possesses flip symmetry $\boldsymbol{P}^x=\prod_{i=1}^N \boldsymbol{\sigma}^x_i$ ($\mathbb{Z}_2$) and 1D translational symmetry $\boldsymbol{T}$. 

We consider the infinite-temperature flip operator as a single-spin flip $\tilde{\boldsymbol{A}}_i=\boldsymbol{\sigma}_i^x$.
To construct the finite-temperature flip operator, we consider the dynamical Lie algebra
\begin{align}
\mathfrak{a}_i^{(0)}&=\boldsymbol{\sigma}^{x}_i,\\
\mathfrak{a}_i^{(1)}&=\frac{1}{\sqrt{2}}(\boldsymbol{\sigma}_{i-1}^{z}+\boldsymbol{\sigma}_{i+1}^{z})\boldsymbol{\sigma}_{i}^{y},\\\mathfrak{a}_i^{(2)}&=\boldsymbol{\sigma}_{i-1}^{z}\boldsymbol{\sigma}_{i}^{x}\boldsymbol{\sigma}_{i+1}^{z}.
\end{align}
The corresponding matrix is given by
\begin{align}
\begin{aligned}
L_{\mathfrak{a}_i}&=2\sqrt{2}J\left(\begin{array}{ccc}
0 & 1 & 0\\
-1 & 0 & -1\\
0 & 1 & 0
\end{array}\right).
\end{aligned}\label{eq:LforclassicalIsing}
\end{align}
The finite-temperature flip operator is
\begin{align}
\begin{aligned}
\boldsymbol{A}_{i}&=\cosh(\frac{\beta J}{2})^{2}\mathfrak{a}_{i}^{(0)}-\frac{i}{\sqrt{2}}\sinh(\beta J)\mathfrak{a}_{i}^{(1)}\\&+\sinh(\frac{\beta J}{2})^{2}\mathfrak{a}_{i}^{(2)}.
\end{aligned}\label{eq:Aiising}
\end{align}
The parent Hamiltonian is constructed according to Eq.~\eqref{eq:Hforclassicalmanybodysystems} as $\boldsymbol{H}=\sum_{i=1}^N\tilde{\gamma}_{i}\boldsymbol{H}_i$ with local terms
\begin{align}
\boldsymbol{H}_{i}&=\frac{1+\eta\boldsymbol{\sigma}_{i}^{z}(\boldsymbol{\sigma}_{i-1}^{z}+\boldsymbol{\sigma}_{i+1}^{z})+\eta^{2}\boldsymbol{\sigma}_{i-1}^{z}\boldsymbol{\sigma}_{i+1}^{z}}{1-\eta^{2}}-\boldsymbol{\sigma}_{i}^{x},
\end{align}
which is supported on sites $i-1$, $i$, and $i+1$. The inverse temperature $\beta$ is given by $\eta:= \tanh(\beta J)$. According to translational symmetry $\boldsymbol{T}$, we choose site-independent kinetic coefficients $\tilde{\gamma}_i = \tilde{\gamma}$.

Since $\boldsymbol{\sigma}_i^z$ commutes with $\boldsymbol{h}^\mathrm{cl}$, we introduce more elaborate flip operators $\tilde{\boldsymbol{A}}_{i1}=(1+\boldsymbol{\sigma}_{i}^{z}\boldsymbol{\sigma}_{i-1}^{z})\boldsymbol{\sigma}_{i}^{x}/2$ and $\tilde{\boldsymbol{A}}_{i2}=(1-\boldsymbol{\sigma}_{i}^{z}\boldsymbol{\sigma}_{i-1}^{z})\boldsymbol{\sigma}_{i}^{x}/2$. The dynamical Lie algebra is isomorphic to the one generated by $\boldsymbol{\sigma}_i^x$. The flip operator $\boldsymbol{A}_{in}$ is obtained from Eq.~\eqref{eq:Aiising} by multiplying it with $(1 \pm \boldsymbol{\sigma}_{i}^{z}\boldsymbol{\sigma}_{i-1}^{z})/2$, and it remains three-body local.
We parameterize the kinetic coefficients as $\tilde{\gamma}_{i1}=\tilde{\gamma}(1+\delta)$ and $\tilde{\gamma}_{i2}=\tilde{\gamma}(1-\delta)$. 
The translational symmetry $\boldsymbol{T}$ ensures identical local terms on all sites, with the same $\tilde{\gamma}$ and $\delta$ applied uniformly.  The parent Hamiltonian for the single-spin flip case is constructed as 
\begin{align}
\begin{aligned}
\boldsymbol{H}&=\sum_{i=1}^{N}\tilde{\gamma}(1+\delta\boldsymbol{\sigma}_{i-1}^{z}\boldsymbol{\sigma}_{i+1}^{z})\boldsymbol{H}_{i}.\label{eq:H_kinetic_Ising_SSF}
\end{aligned}
\end{align}
The free parameter $\delta$ can be exploited to enhance the spectral gap. We remark that Glauber dynamics is recovered by choosing $\delta = -\eta^2$, which gives the normalized gap 
\begin{align}
    \Delta_{\mathrm{Glauber}}=\frac{2(1-\eta)^{2}}{N(1+\eta^{2})}.
\end{align}
For this choice, the coefficient of $\boldsymbol{\sigma}^z_{i-1}\boldsymbol{\sigma}^z_{i+1}$ in Eq.~\eqref{eq:H_kinetic_Ising_SSF} vanishes. The Hamiltonian maps to free Majorana fermions, and the spectrum is exactly solvable.

\subsection{1D double-spin flip kinetic Ising model}
The spectral gap can be further enhanced by combining the single-spin and double-spin flips. In the following, we exploit the dynamical Lie algebra to construct the dynamics for double-spin flips.  As in the single-spin flip case, projection operators are employed to construct more elaborate flip operators. Using the symmetry argument, the kinetic coefficients of these flip operators should respect the symmetry group of the thermal state. 

We consider the infinite-temperature flip operator as a double-spin flip $\tilde{\boldsymbol{A}}_{i,i+1}=\boldsymbol{\sigma}_{i}^x \boldsymbol{\sigma}_{i+1}^x$,
and construct the dynamical Lie algebra with 
\begin{align}
\mathfrak{a}_{i,i+1}^{(0)}&=\boldsymbol{\sigma}_{i}^{x}\boldsymbol{\sigma}_{i+1}^{x},\\\mathfrak{a}_{i,i+1}^{(1)}&=\frac{1}{\sqrt{2}}(\boldsymbol{\sigma}_{i-1}^{z}\boldsymbol{\sigma}_{i}^{y}\boldsymbol{\sigma}_{i+1}^{x}+\boldsymbol{\sigma}_{i}^{x}\boldsymbol{\sigma}_{i+1}^{y}\boldsymbol{\sigma}_{i+2}^{z}),\\\mathfrak{a}_{i,i+1}^{(2)}&=\boldsymbol{\sigma}_{i-1}^{z}\boldsymbol{\sigma}_{i}^{y}\boldsymbol{\sigma}_{i+1}^{y}\boldsymbol{\sigma}_{i+2}^{z},
\end{align}
while the structure matrix $L_{\mathfrak{a}_{i,i+1}}$ is still given by Eq.~\eqref{eq:LforclassicalIsing}. Therefore, the finite-temperature flip operator is 
\begin{align}
\begin{aligned}
\boldsymbol{A}_{i,i+1}&=\cosh(\frac{\beta J}{2})^{2}\mathfrak{a}_{i,i+1}^{(0)}-\frac{i}{\sqrt{2}}\sinh(\beta J)\mathfrak{a}_{i,i+1}^{(1)}\\&+\sinh(\frac{\beta J}{2})^{2}\mathfrak{a}_{i,i+1}^{(2)}.
\end{aligned}
\end{align}
The parent Hamiltonian is $\boldsymbol{H}=\sum_{i=1}^N \tilde{\gamma}_{i,i+1}\boldsymbol{H}_{i,i+1}$ with the local term 
\begin{align}
\begin{aligned}
    \boldsymbol{H}_{i,i+1}&=\frac{1}{1-\eta^{2}}+\frac{\eta}{1-\eta^{2}}(\boldsymbol{\sigma}_{i-1}^{z}\boldsymbol{\sigma}_{i}^{z}+\boldsymbol{\sigma}_{i+1}^{z}\boldsymbol{\sigma}_{i+2}^{z})\\&+\frac{\eta^{2}}{1-\eta^{2}}\boldsymbol{\sigma}_{i-1}^{z}\boldsymbol{\sigma}_{i}^{z}\boldsymbol{\sigma}_{i+1}^{z}\boldsymbol{\sigma}_{i+2}^{z}-\boldsymbol{\sigma}_{i}^{x}\boldsymbol{\sigma}_{i+1}^{x},
\end{aligned}
\end{align}
which is supported on sites $i-1$, $i$, $i+1$, and $i+2$ with the same $\eta = \tanh(\beta J)$ as above.
More elaborate double-spin flips can be constructed by multiplying $\boldsymbol{A}_{i,i+1}$ with the projector factors $(1\pm\boldsymbol{\sigma}^z_{i-1}\boldsymbol{\sigma}^z_{i+2})/2$, $(1\pm\boldsymbol{\sigma}^z_{i}\boldsymbol{\sigma}^z_{i+1})/2$, and $(1\pm\boldsymbol{\sigma}^z_{i-1}\boldsymbol{\sigma}^z_{i}\boldsymbol{\sigma}^z_{i+1}\boldsymbol{\sigma}^z_{i+2})/2$.
The parent Hamiltonian for the general double-spin flips is constructed as
\begin{align}
    \begin{aligned}
\boldsymbol{H}&=\sum_{i=1}^{N}\tilde{\gamma}_{2}(1+\delta_{1}\boldsymbol{\sigma}_{i}^{z}\boldsymbol{\sigma}_{i+1}^{z}+\delta_{2}\boldsymbol{\sigma}_{i-1}^{z}\boldsymbol{\sigma}_{i+2}^{z}\\&+\delta_{3}\boldsymbol{\sigma}_{i}^{z}\boldsymbol{\sigma}_{i+1}^{z}\boldsymbol{\sigma}_{i-1}^{z}\boldsymbol{\sigma}_{i+2}^{z})\boldsymbol{H}_{i,i+1},
    \end{aligned}\label{eq:H_kinetic_Ising_DSF}
\end{align}
with kinetic coefficients $\tilde{\gamma}_2$, $\delta_1$, $\delta_2$, and $\delta_3$. Here the translational symmetry is already imposed.

\begin{figure}
    \centering
    \includegraphics[width=\columnwidth]{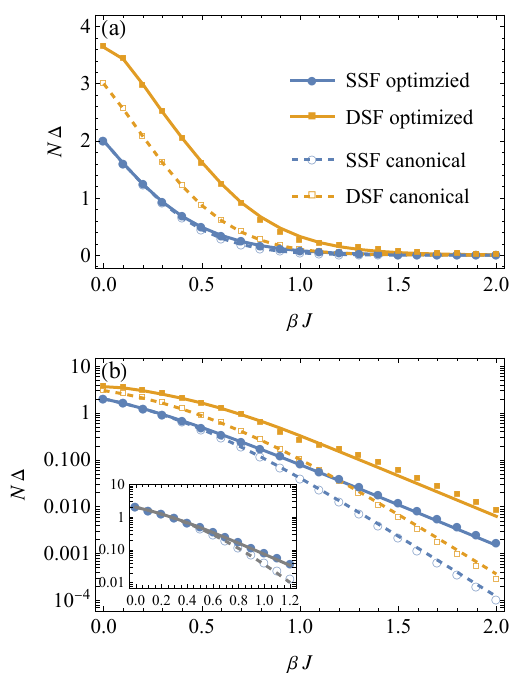}
    \caption{The normalized gap $N\Delta$ as a function of inverse temperature $\beta$ for the 1D kinetic Ising model, 
    shown on (a) linear and (b) logarithmic scales. 
    Circles correspond to the single-spin flip case, and squares to the case combining the single- and the double-spin flips.
Dashed and solid curves show exact-diagonalization results for the canonical and optimized kinetic coefficients at $N=10$, while open and filled markers show the corresponding DMRG results at $N=40$.
In the inset of panel (b), dashed and solid gray curves show the SDP-certified gaps for the single-spin flip case with the canonical and optimized kinetic coefficients, respectively. The markers show the DMRG results for $N=40$.}
    \label{fig:optimalSSFandDSF}
\end{figure}

We evaluate and optimize the spectral gap of the parent Hamiltonians in Eqs.~\eqref{eq:H_kinetic_Ising_SSF} and \eqref{eq:H_kinetic_Ising_DSF}. Figure~\ref{fig:optimalSSFandDSF}(a) and (b) show the normalized gap on linear and logarithmic scales, respectively. 
We compare the canonical (dashed curves and open markers) and optimized (solid curves and filled markers) choices of kinetic coefficients for both the single-spin flip case (circles) and the case combining single- and double-spin flips (squares). 
In the canonical choice, we set $\delta=0$ and keep only $\tilde{\gamma}$ nonzero for the single-spin flip case.  
For the case combining single-spin flip and double-spin flips,
we set $\delta=\delta_1=\delta_2=\delta_3=0$
and choose the two nonzero kinetic coefficients equal, $\tilde{\gamma}=\tilde{\gamma}_2$. 
In the optimized choice, we optimize the kinetic coefficients at system size $N=10$ to enhance the spectral gap.  
For the single-spin flip case, only $\tilde{\gamma}$ and $\delta$ are optimized. 
For the case combining single-spin and double-spin flips, we optimize all kinetic coefficients $\tilde{\gamma}$, $\tilde{\gamma}_2$, $\delta$, $\delta_1$, $\delta_2$, and $\delta_3$. 
The optimization is carried out at system size $N=10$. We compare the exact diagonalization results for $N=10$ (curves) with the DMRG results for $N=40$ (markers), which agree well.
For the single-spin flip case, the certified lower bound (gray curves) of the spectral gap is shown in the inset of Fig.~\ref{fig:optimalSSFandDSF}(b), which is obtained  by formulating the spectral gap certification as the SDP on six neighboring sites \cite{rai_hierarchy_2024}.

While the canonical gap is large at high temperatures, the enhancement becomes more pronounced as the temperature decreases. Optimizing only the single–spin flip leads to appreciable enhancement (lower blue circles and curves), and optimizing both single- and double-spin flips simultaneously leads to a larger enhancement in the spectral gap (upper orange squares and curves). Optimization performed at a smaller system size $N=10$ remains effective for larger system size $N=40$, giving up to a tenfold increase in the gap for $\beta J \ge 1$. Moreover, the spectral gap can be certified only up to $\beta J \le 1.3$ for the canonical kinetic coefficients, while for the optimized kinetic coefficients it can be certified at least up to $\beta J=2$. This implies that the spectral gap is easier to certify for the optimized kinetic coefficients.

\subsection{1D Transverse-field Ising model}
We now apply the framework to the transverse-field Ising model \cite{Sachdev2015,Dziarmaga2005}. The Hamiltonian of the 1D transverse-field Ising model is $\boldsymbol{h}=\sum_{i=1}^N \boldsymbol{h}_i $, where the local term is 
\begin{align}
\boldsymbol{h}_{i}=-J\left(\boldsymbol{\sigma}_{i}^{z}\boldsymbol{\sigma}_{i+1}^{z}+g\boldsymbol{\sigma}_{i}^{x}\right).
\end{align}
We impose periodic boundary conditions $\boldsymbol{\sigma}_{N+1}^z=\boldsymbol{\sigma}_1^z$.
The classical Ising Hamiltonian is recovered by setting $g= 0$.

We first construct the dynamical Lie algebra from the infinite-temperature Lindblad operators, chosen as $\tilde{\boldsymbol{A}}_{i1}=\boldsymbol{\sigma}_i^x$, $\tilde{\boldsymbol{A}}_{i2}=\boldsymbol{\sigma}_i^y$, and $\tilde{\boldsymbol{A}}_{i3}=\boldsymbol{\sigma}_i^z$. 
The dynamical Lie algebra is generated by these local operators through successive commutators with the Hamiltonian $\boldsymbol{h}$. Unlike the classical case, nested commutators generate operators with increasingly larger spatial support, eventually spanning all operators consistent with the symmetry and locality constraints of the Hamiltonian.
The finite-temperature Lindblad operators are then obtained from Eq.~\eqref{eq:A_{i}dynamicalLie} as $\boldsymbol{A}_{in}=\mathfrak{a}_{in}^\dagger  e^{\frac{i\beta}{4}L_{\mathfrak{a}_{in}}}E_0$,
where $\mathfrak{a}_{in}^\dagger$ denotes the basis of the dynamical Lie algebra generated by $\tilde{\boldsymbol{A}}_{in}$, and $L_{\mathfrak{a}_{in}}$ is the corresponding matrix. This operator growth from local operators to system-size operators reflects both the nearest-neighbor coupling structure and the algebraic closure of the dynamical Lie algebra of the transverse-field Ising model.
We impose translational symmetry on the kinetic coefficients $\tilde{\gamma}_{in}=\tilde{\gamma}_n$, and the dissipation term is simplified to $\boldsymbol{R}=\sum_{n=1}^{3}\tilde{\gamma}_{n}\sum_{i=1}^{N}\boldsymbol{A}_{in}^{\dagger}\boldsymbol{A}_{in}$.
We compute the dynamical Lie algebra $\mathfrak{r}$ generated by $\boldsymbol{R}$, and construct $\boldsymbol{K}$ and $\boldsymbol{N}$ with the corresponding matrix $L_{\mathfrak{r}}$.

\begin{figure}
    \centering
    \includegraphics[width=\linewidth]{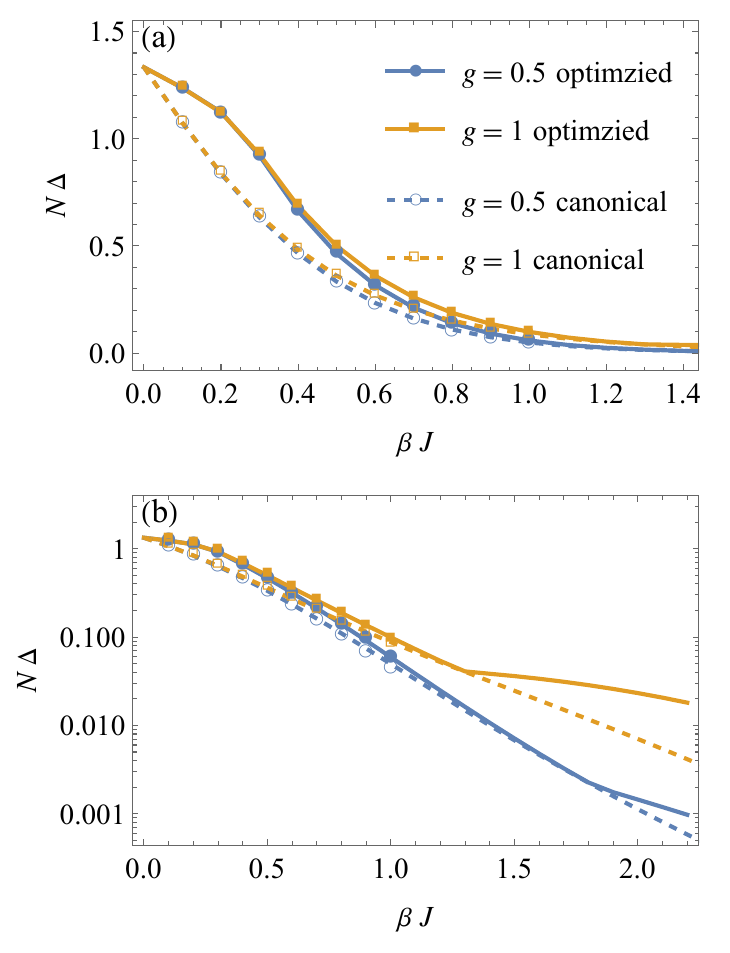}
    \caption{The normalized gap $N\Delta$ as a function of $\beta J$ for the 1D transverse-field Ising model, shown on (a) linear and (b) logarithmic scales.
Circles correspond to $g=0.5$, and squares to $g=1$.
The spectral gaps in both cases are obtained from the parent Hamiltonian in the thermofield-double space.
Dashed and solid curves represent exact-diagonalization results for the canonical and optimized kinetic coefficients at $N=5$, while open and filled markers denote the corresponding DMRG results at $N=20$.}
    \label{fig:transverse_Ising_fig}
\end{figure}

Figure~\ref{fig:transverse_Ising_fig} shows the spectral gap for the transverse-field Ising model as a function of the inverse temperature $\beta J$ for transverse fields $g=0.5$ (circles) and $g=1$ (squares), and for system sizes $N=5$ and $N=20$. We compare the normalized gap $N\Delta$ for the canonical (dashed curves and empty markers) and optimized (solid curves and filled markers) kinetic coefficients on (a) linear and (b) logarithmic scales. To prepare the quantum thermal state, we construct the
parent Hamiltonian $\mathcal{H}$  in the thermofield-double space with $2N$ sites. For the small system size $N=5$, we solve Eq.~\eqref{eq:equation_to_solve_N} for $\boldsymbol{N}$ to construct the parent Hamiltonian and obtain the spectral gap by exact diagonalization (curves). For the larger system size $N=20$, 
we use the dynamical Lie algebras (Eqs.~\eqref{eq:Atildetquantum} and~\eqref{eq:Ntermliealgebra}) to construct the parent Hamiltonian with truncation orders of $l_{\mathfrak{a}}=10$ and $l_{\mathfrak{r}}=20$ for $\beta J\leq1$. At lower temperatures, larger truncation orders would be required, which introduces more nonlocal operators; the increased nonlocality of the parent Hamiltonian makes the DMRG less efficient.
With the chosen truncation, the errors in the coefficients of the parent Hamiltonian remain below $10^{-6}$. 
We retain only the local terms with the coefficients larger than $10^{-6}$, which yields an error smaller than $3\times 10^{-5}$ in the ground-state energy. 
The spectral gap for $N=20$ is then obtained using DMRG for $\beta J\leq 1$ (markers). 

\section{Conclusions}\label{sec.conclusion}

Determining the transient and steady states of open
quantum systems is a central problem in the development of quantum technologies. In this paper, we focused
on describing and engineering the transient dynamics to
favor fast relaxation toward thermal states. The results obtained here are relevant to several areas of quantum technologies, including quantum thermodynamics, quantum simulation, and quantum algorithms.
Our contributions address three main aspects: (i) construction of Liouvillians and Lindbladians preparing target thermal states; (ii) variational and SDP approaches to approximate and certify the minimum decay rate; and (iii) optimization of kinetic coefficients to enhance the spectral gap and boost thermalization.

A central result of this work is a general transformation between finite- and infinite-temperature Lindbladians.
This transformation relies on the quantum detailed balance condition, 
which implies that these Lindbladians correspond to frustration-free parent Hamiltonians in the thermofield-double space. 
In the quantum regime, noncommutativity renders the finite-temperature Lindbladians quasilocal, 
but their locality can be improved by appropriately filtering the infinite-temperature Lindblad operators, a technique recently exploited in the first quantum Gibbs sampler for noncommuting Hamiltonians \cite{chen_efficient_2025}. The dynamical Lie algebra approach remains applicable in this setting. 
Moreover, our construction of relaxation dynamics and optimization of the spectral gap also apply to classical systems, where the Liouvillian remains strictly local.

The transient dynamics of open systems are described by the spectrum of the Lindbladian, which quickly becomes inaccessible for large systems.
In this regard, we provide upper and lower bounds for the minimum decay rate using variational methods and semidefinite programming techniques, respectively.
To the best of our knowledge, the SDP-based lower bound provides a systematic way to certify this critical quantity in open-system dynamics.

By exploiting freedom in the choice of kinetic coefficients, we demonstrate the optimization of these coefficients to enhance the spectral gap, which reduces thermalization timescales. According to Weyl's inequality \cite{weyl_asymptotische_1912}, the spectral gap depends concavely on convex combinations of Lindbladians that share the same equilibrium state. Therefore, the optimal Lindbladian is expected to respect the symmetries of the equilibrium state, thereby reducing the number of free parameters in the kinetic coefficients. 

In future research, 
applying the framework to quantum many-body systems with noncommuting Hamiltonians requires efficient representations (e.g. tensor-network \cite{cirac_matrix_2021}, neural quantum states \cite{carleo_solving_2017}) for quasilocal Lindbladians or parent Hamiltonians. We leave this open challenge for future study.
Also, utilizing the SDP approach \cite{cruz_preparation_2022,kull_lower_2024,robichon_bootstrapping_2024,mortimer_certifying_2025} could certify lower bounds on spectral gaps for larger systems. 
Combining variational upper bounds with SDP lower bounds will improve the optimization of Lindblanians and allow extensions to larger systems
with more complex kinetic rules (e.g., multiple spin-flip), more spatial dimensions (e.g., 2D kinetic Ising model).  
These bounds further enable precise characterization of the dynamical critical exponent of kinetic Ising models \cite{haake_universality_1980}.
A compelling direction for future work is to invert the problem: rather than maximizing the minimum decay rate to boost thermalization, one could minimize the maximum decay rate to reduce environmental disturbance. This approach involves finding an upper bound for the maximum decay rate in open many-body systems, as recently demonstrated in Ref.~\cite{mok2024universal}.

Additionally, incorporating insights from quantum thermodynamics will deepen our understanding of quantum work and heat in the context of open many-body systems. 
Optimizing control in these systems would benefit from combining static (choice of kinetic coefficients) and protocol optimization (schedule for changing the control parameters), 
thereby integrating spectral gap enhancement with non-equilibrium thermodynamic length ~\cite{rezakhani_quantum_2009,sivak_thermodynamic_2012,PhysRevLett.119.050601,scandi_thermodynamic_2019,chen_extrapolating_2021,chen_speeding_2022,blaber_optimal_2023}. 
Our approach enables more effective design, manipulation, and optimization of quantum many-body systems.

\acknowledgements
We thank Antonio Acín,  Hui Dong and Tao Shi for useful
discussions. J.F.C. and J.T. also thank the helpful discussion with András Gilyén and Michael J. Kastoryano on the CKG algorithm during Qtech'25 Conference in Budapest.

J.F.C., K.S.R., and J.T. acknowledge the support received from the European Union's Horizon Europe research and innovation programme through the ERC StG FINE-TEA-SQUAD (Grant No.~101040729). P.E. and J.T. acknowledge the support received by the Dutch National Growth Fund
(NGF), as part of the Quantum Delta NL programme. 
P.E. acknowledges the support received through the NWO-Quantum Technology programme (Grant No.~NGF.1623.23.006) and funding by the Carl-Zeiss-Stiftung (CZS Center QPhoton).
This publication is part of the `Quantum Inspire - the Dutch Quantum Computer in the Cloud' project (with project number [NWA.1292.19.194]) of the NWA research program `Research on Routes by Consortia (ORC)', which is funded by the Netherlands Organization for Scientific Research (NWO).
Parts of this work were performed by using the compute resources from the Academic Leiden Interdisciplinary Cluster Environment (ALICE) provided by Leiden University.
The views and opinions expressed here are solely those of the authors and do not necessarily reflect those of the
funding institutions. Neither of the funding institutions can be held responsible for them.

D.F. acknowledges financial support from PNRR MUR Project No. PE0000023-NQSTI.
M.P, P.G., and M.L. acknowledges support from:
European Research Council AdG NOQIA; MCIN/AEI (PGC2018-0910.13039/501100011033, CEX2019-000910-S/10.13039/501100011033, Plan National FIDEUA PID2019-106901GB-I00, 
Plan National STAMEENA PID2022-139099NB, I00, project funded by MCIN/AEI/10.13039/501100011033 and by the “European Union NextGenerationEU/PRTR" (PRTR-C17.I1), FPI); 
QUANTERA MAQS PCI2019-111828-2; QUANTERA DYNAMITE PCI2022-132919, QuantERA II Programme co-funded by European Union’s Horizon 2020 program under Grant Agreement No 101017733; Ministry for Digital Transformation and of Civil Service of the Spanish Government through the QUANTUM ENIA project call - Quantum Spain project, and by the European Union through the Recovery, Transformation and Resilience Plan - NextGenerationEU within the framework of the Digital Spain 2026 Agenda; 
Fundació Cellex; Fundació Mir-Puig; Generalitat de Catalunya (European Social Fund FEDER and CERCA program, AGAUR Grant No. 2021 SGR 01452, QuantumCAT U16-011424, co-funded by ERDF Operational Program of Catalonia 2014-2020); Barcelona Supercomputing Center MareNostrum (FI-2023-3-0024);  Funded by the European Union.

Views and opinions expressed are however those of the author(s) only and do not necessarily reflect those of the European Union, 
European Commission, 
European Climate, Infrastructure and Environment Executive Agency (CINEA), or any other granting authority.  Neither the European Union nor any granting authority can be held responsible for them 
(HORIZON-CL4-2022-QUANTUM-02-SGA  
PASQuanS2.1, 101113690, EU Horizon 2020 FET-OPEN OPTOlogic,
Grant No 899794),  EU Horizon Europe Program (This project has received funding from the European Union’s Horizon Europe research and innovation program under grant agreement No 101080086 NeQSTGrant Agreement 101080086 — NeQST);  ICFO Internal “QuantumGaudi” project;  European Union’s Horizon 2020 program under the Marie Sklodowska-Curie grant agreement No 847648; “La Caixa” Junior Leaders fellowships, La Caixa” Foundation (ID 100010434): CF/BQ/PR23/11980043. 
\appendix

\section{Superoperator}\label{Appedix_A_superoperator}
It is convenient to vectorize the density matrix so that superoperators become ordinary matrices acting on vectors. Explicitly, we write the density matrix as
\begin{align}
\boldsymbol{\rho} = 
\begin{pmatrix}
\rho_{11} & \rho_{12} & \cdots & \rho_{1d} \\
\rho_{21} & \rho_{22} & \cdots & \rho_{2d} \\
\vdots    & \vdots    & \ddots & \vdots    \\
\rho_{d1} & \rho_{d2} & \cdots & \rho_{dd}
\end{pmatrix}.
\end{align}
By stacking all rows, $\boldsymbol{\rho}$ is reshaped into a column vector
\begin{align}
\ket{\boldsymbol{\rho}}=\begin{pmatrix}\rho_{11},\cdots,\rho_{1d},\rho_{21},\cdots,\rho_{2d},\cdots,\rho_{d1},\cdots,\rho_{dd}\end{pmatrix}^{\top},
\end{align}
which we denote by a ket and refer to as the vectorization of $\boldsymbol{\rho}$.  

In this representation, the action of a superoperator on $\boldsymbol{\rho}$ can be written as
\begin{align}
\mathcal{D}[\boldsymbol{\rho}] \;\rightarrow\; \mathcal{D}\ket{\boldsymbol{\rho}},
\end{align}
where $\mathcal{D}$ is now a $d^{2}\times d^{2}$ matrix.  
For the Lindbladian in Eq.~\eqref{eq:Dissipator_detailed_balance_condition}, the matrix representation reads
\begin{align}
\begin{aligned}
    \mathcal{D}=&-(\frac{1}{2}\boldsymbol{R}+i\,\boldsymbol{K})\otimes\boldsymbol{I}-\,\boldsymbol{I}\otimes(\frac{1}{2}\boldsymbol{R}^{\top}-i\boldsymbol{K}^{\top})\\&+\sum_{n}\tilde{\gamma}_{n}\boldsymbol{A}_{n}\otimes\boldsymbol{A}_{n}^{*}.
\end{aligned}
\end{align}

Similarly, the parent Hamiltonian defined in Eq.~\eqref{eq:parent_H_quantum} takes the form
\begin{align}
\mathcal{H}=-(\boldsymbol{\rho}_{\beta}^{-1/4}\otimes\boldsymbol{\rho}_{\beta}^{-1/4\top})\mathcal{D}(\boldsymbol{\rho}_{\beta}^{1/4}\otimes\boldsymbol{\rho}_{\beta}^{1/4\top}),
\end{align}
whose ground state corresponds to the thermofield-double state $\ket{\boldsymbol{\rho}_{\beta}^{1/2}}$. Equation~\eqref{eq:parent_H_thermofielddouble} becomes
\begin{align}
\mathcal{H}=&\frac{1}{2}\boldsymbol{N}\otimes\boldsymbol{I}+\frac{1}{2}\boldsymbol{I}\otimes\boldsymbol{N}^{\top}-\sum_{n}\tilde{\gamma}_{n}\tilde{\boldsymbol{A}}_{n}\otimes\tilde{\boldsymbol{A}}_{n}^{*}.\label{eq:parent_H_thermofielddoublevectorized}
\end{align}
The full spectrum of the dynamics can thus be obtained by diagonalizing the matrices $\mathcal{D}$ or $\mathcal{H}$.

The purified state $\ket{\boldsymbol{\rho}^{1/2}}$ can be represented as a matrix product state, while the superoperators $\mathcal{D}$ and $\mathcal{H}$ are naturally expressed as matrix product operators.
Within this framework, the action of $\mathcal{H}$ corresponds to a matrix product operator acting on a matrix product state in the doubled Hilbert space.
This tensor-network formulation enables numerical methods such as DMRG to access the spectrum and steady states of Lindbladian dynamics.

\section{Derivation to Eqs.~\eqref{eq:equation_to_solve_K} and~\eqref{eq:Kenergydomain}}\label{apppendix D:derivation to G}

We use Eq.~\eqref{eq:D_quantum_detailed_balance} to obtain 
\begin{align}
\begin{aligned}
    &(-\frac{1}{2}\boldsymbol{R}+i\boldsymbol{K})\bullet+\bullet(-\frac{1}{2}\boldsymbol{R}-i\boldsymbol{K})+\sum_{n}\tilde{\gamma}_{n}\boldsymbol{A}_{n}^{\dagger}\bullet\boldsymbol{A}_{n}\\
    =&\sum_{n}\tilde{\gamma}_{n}\boldsymbol{\rho}_{\beta}^{-1/2}\boldsymbol{A}_{n}\boldsymbol{\rho}_{\beta}^{1/2}\bullet\boldsymbol{\rho}_{\beta}^{1/2}\boldsymbol{A}_{n}^{\dagger}\boldsymbol{\rho}_{\beta}^{-1/2}\\
    &+(-\frac{1}{2}\boldsymbol{\rho}_{\beta}^{-1/2}\boldsymbol{R}\boldsymbol{\rho}_{\beta}^{1/2}-i\boldsymbol{\rho}_{\beta}^{-1/2}\boldsymbol{K}\boldsymbol{\rho}_{\beta}^{1/2})\bullet\\
    &+\bullet(-\frac{1}{2}\boldsymbol{\rho}_{\beta}^{1/2}\boldsymbol{R}\boldsymbol{\rho}_{\beta}^{-1/2}+i\boldsymbol{\rho}_{\beta}^{1/2}\boldsymbol{K}\boldsymbol{\rho}_{\beta}^{-1/2}).
\end{aligned}
\end{align}
Since the jump terms are chosen to satisfy $\sum_{n}\tilde{\gamma}_{n}\boldsymbol{A}_{n}^{\dagger}\bullet\boldsymbol{A}_{n}=\sum_{n}\tilde{\gamma}_{n}\boldsymbol{\rho}_{\beta}^{-1/2}\boldsymbol{A}_{n}\boldsymbol{\rho}_{\beta}^{1/2}\bullet\boldsymbol{\rho}_{\beta}^{1/2}\boldsymbol{A}_{n}^{\dagger}\boldsymbol{\rho}_{\beta}^{-1/2}$, we only need to impose
\begin{align}
    -\frac{1}{2}\boldsymbol{R}+i\boldsymbol{K}&=-\frac{1}{2}\boldsymbol{\rho}_{\beta}^{-1/2}\boldsymbol{R}\boldsymbol{\rho}_{\beta}^{1/2}-i\boldsymbol{\rho}_{\beta}^{-1/2}\boldsymbol{K}\boldsymbol{\rho}_{\beta}^{1/2},\label{eq:R+iK}
\end{align}
which gives Eq.~\eqref{eq:equation_to_solve_K}.

By transforming Eq.~\eqref{eq:R+iK} into the energy domain, one obtains the expression
\begin{align}
\sum_{\nu}\!\left(-\tfrac{1}{2}\boldsymbol{R}_{\nu}+i\boldsymbol{K}_{\nu}\right)
=\sum_{\nu} e^{\beta\nu/2}\!\left(-\tfrac{1}{2}\boldsymbol{R}_{\nu}-i\boldsymbol{K}_{\nu}\right),
\end{align}
from which Eq.~\eqref{eq:Kenergydomain} follows immediately.

\section{Krylov space expansion for operator dynamics}\label{sec:Krylov space expansion for operator dynamics}

To simulate the Heisenberg evolution of an operator $\boldsymbol{A}(t) = e^{i\boldsymbol{h}t}\boldsymbol{A}\,e^{-i\boldsymbol{h}t}$, 
we represent the dynamics in terms of the Liouvillian superoperator
\begin{align}
\mathcal{L}_{\boldsymbol{h}}[\bullet] = i[\boldsymbol{h},\bullet],
\end{align}
which generates a Hermitian operator with respect to the normalized Hilbert--Schmidt inner product. Here we take $\boldsymbol{A}$ to be traceless, since the constant component can be subtracted.

Rather than working in the full operator space of dimension $d^2$, 
we approximate the dynamics in the Krylov subspace generated by $\mathcal{L}_{\boldsymbol{h}}$:
\begin{align}
\mathscr{K}(\mathcal{L}_{\boldsymbol{h}},\boldsymbol{A})=\mathrm{span}\{\boldsymbol{A},\mathcal{L}_{\boldsymbol{h}}[\boldsymbol{A}],\ldots,\mathcal{L}_{\boldsymbol{h}}^{l-1}[\boldsymbol{A}],\ldots\}.
\end{align}
In practice, we truncate this basis to a finite dimension $l_{\mathfrak{a}}$.
An orthonormal basis is denoted
\begin{align}
\mathfrak{a} ^\dagger = \big(\mathfrak{a}^{(0)}, \mathfrak{a}^{(1)}, \ldots, \mathfrak{a}^{(l_{\mathfrak{a}}-1)}\big),
\end{align}
with $\mathfrak{a}^{(0)} = \boldsymbol{A}/\|\boldsymbol{A}\|$ and $\|\boldsymbol{A}\|=\sqrt{\left\langle \boldsymbol{A},\boldsymbol{A}\right\rangle }$. The Lanczos recursion generates successive basis elements via
\begin{align}
\mathcal{L}_{\boldsymbol{h}}[\mathfrak{a}^{(l)}]=\mathfrak{a}^{(l-1)}L_{\mathfrak{a}}^{(l-1,l)}+\mathfrak{a}^{(l+1)}L_{\mathfrak{a}}^{(l+1,l)},
\end{align}
where $L_{\mathfrak{a}}$ is a tridiagonal skew-symmetric matrix. 

The time evolution is then approximated in the truncated space as
\begin{align}
\boldsymbol{A}(t)\approx\|\boldsymbol{A}\|\mathfrak{a}^{\dagger}e^{L_{\mathfrak{a}}t}E_{0},
\end{align}
with $E_0=(1,0,\ldots,0)^\top$.
This Krylov–Lanczos approach offers a systematic framework for constructing Lindbladians and parent Hamiltonians in large systems.

\section{Remarks on CKG algorithm}\label{sec:Remarks on CKG algorithm}

We show that the operator Fourier transform  underlying the CKG algorithm \cite{chen_efficient_2025} can be combined with the dynamical Lie algebra, leading to improved locality of the parent Hamiltonians. The operator Fourier transform produces energy-filtered operators $\tilde{\boldsymbol{A}}_{i,\pm}$, and applying a Gaussian window enhances the locality of the resulting finite-temperature Lindblad operators. A subsequent singularity transform gives $\boldsymbol{A}_{i,\pm}$, from which the dissipation term $\boldsymbol{R}$ follows. The dynamical Lie algebra generated by $\boldsymbol{R}$ then determines $\boldsymbol{K}$ and $\boldsymbol{N}$. Translational symmetry can be exploited in evaluating the dynamical Lie algebra generated by $\boldsymbol{R}$.

To improve the locality of the parent Hamiltonian, we choose a pair of Lindblad operators $\tilde{\boldsymbol{A}}_{i,\pm}$ with an energy filter
\begin{align}
    \tilde{\boldsymbol{A}}_{i,\pm}=\hat{\boldsymbol{B}}_{i}(\pm\tilde{\omega}),
    \end{align}
    where the operator Fourier transform is defined as \cite{chen_efficient_2025}
    \begin{align}
    \hat{\boldsymbol{B}}_{i}(\tilde{\omega})&=\frac{1}{\sqrt{2\pi}}\int_{-\infty}^{\infty}e^{-i\tilde{\omega}t}f(t)\boldsymbol{B}_{i}(t)dt.
    \label{eq:operator_Fourier}
    \end{align}
    We choose $\boldsymbol{B}_i$ to be local operators, for example, the Pauli operator $\boldsymbol{\sigma}^x_i$ acting on a single site $i$, and use the filter $f(t)=e^{-\sigma_E ^2t^2}\sqrt{\sigma_E\sqrt{2/\pi}}$ to select the transitions with energy change around $\tilde{\omega}$. The pair $\tilde{\boldsymbol{A}}_{i,\pm}$ is assigned the same rate $\tilde{\gamma}_{i,+}=\tilde{\gamma}_{i,-}=\tilde{\gamma}_i$ to ensure the detailed balance condition [Eq.~\eqref{eq:infinitetemperature_jump_detailed_balance}] for the jump term. 

Within the framework of the dynamical Lie algebra, we write 
\begin{align}
    \boldsymbol{B}_i (t)=\mathfrak{b}_i^\dagger e^{L_{\mathfrak{b}_i}t}E_0.
\end{align}
The operator Fourier transform then becomes 
\begin{align}
\begin{aligned}
\hat{\boldsymbol{B}}_{i}(\tilde{\omega})&=\left(\frac{\sigma_{E}^{2}}{2\pi^{3}}\right)^{\frac{1}{4}}\int_{-\infty}^{\infty}e^{-\sigma_{E}^{2}t^{2}}e^{-i\tilde{\omega}t}\boldsymbol{B}_{i}(t)dt
\\&=\frac{e^{-\tilde{\omega}^{2}/(4\sigma_{E}^{2})}}{(2\pi\sigma_{E}^{2})^{1/4}}\mathfrak{b}_{i}^{\dagger}e^{\frac{L_{\mathfrak{b}_{i}}^{2}-2i\tilde{\omega}L_{\mathfrak{b}_{i}}}{4\sigma_{E}^{2}}}E_{0}.
\end{aligned}
\end{align}
The Lindblad operators at finite temperature become
\begin{align}
\begin{aligned}
    \boldsymbol{A}_{i,\pm}&=\frac{1}{\sqrt{2\pi}}\int_{-\infty}^{\infty}e^{\mp i\tilde{\omega}t}f(t)e^{-\frac{\beta}{4}\boldsymbol{h}}\boldsymbol{B}_{i}(t)e^{\frac{\beta}{4}\boldsymbol{h}}dt\\&=\frac{1}{\sqrt{2\pi}}\int_{-\infty}^{\infty}e^{\mp i\tilde{\omega}t}f(t)\boldsymbol{B}_{i}(t+\frac{i\beta}{4})dt\\&=\frac{1}{\sqrt{2\pi}}\int_{-\infty}^{\infty}e^{\mp i\tilde{\omega}(t-\frac{i\beta}{4})}f(t-\frac{i\beta}{4})\boldsymbol{B}_{i}(t)dt\\&=\frac{e^{-\tilde{\omega}^{2}/(4\sigma_{E}^{2})}}{(2\pi\sigma_{E}^{2})^{1/4}}\mathfrak{b}_{i}^{\dagger}e^{\frac{i\beta L_{\mathfrak{b}_{i}}}{4}+\frac{L_{\mathfrak{b}_{i}}^{2}-2i\tilde{\omega}L_{\mathfrak{b}_{i}}}{4\sigma_{E}^{2}}}E_{0},
\end{aligned}
\end{align}
which are used to construct the dissipation term $\boldsymbol{R}$.
The parent Hamiltonian is then given by 
\begin{align}
\mathcal{H}[\bullet]&=\frac{1}{2}\{\boldsymbol{N},\bullet\}-\sum_{i=1}^N\sum_{k=\pm}\tilde{\gamma}_{i}\tilde{\boldsymbol{A}}_{ik}\bullet\tilde{\boldsymbol{A}}_{ik}^{\dagger},\label{eq:parentHCKGapair}
\end{align}
where $\boldsymbol{N}$ is obtained by solving the dynamical Lie algebra from $\boldsymbol{R}$. Equation~\eqref{eq:parentHCKGapair} defines the parent Hamiltonian associated with a pair of Lindblad operators. The Gaussian-weighted parent Hamiltonians and Lindbladians of Ref.~\cite{chen_efficient_2025} are recovered by introducing a Gaussian weight over the transition energy $\tilde{\omega}$.

\bibliographystyle{apsrev4-1}
\bibliography{c2q_hamiltonian,paper/bibfile}
\end{document}